\documentclass[twocolumn]{aastex631}
\usepackage[all]{hypcap} 
\usepackage[utf8]{inputenc}
\usepackage{amsmath}
\usepackage{booktabs}
\newcommand{\rate}{\tilde{\mathcal{R}}}

\color{black}

\begin{document}

\title{Studying Binary Formation under Dynamical Friction Using Hill's Problem}
\shorttitle{DYNAMICAL FRICTION FORMING BINARY BLACK HOLES}
\shortauthors{DODICI \& TREMAINE}

\author[0000-0002-3352-9272]{Mark Dodici}
\affiliation{Department of Astronomy \& Astrophysics, University of Toronto, Toronto, Ontario, Canada}
\affiliation{Canadian Institute for Theoretical Astrophysics, Toronto, Ontario, Canada}
\email{mark.dodici@astro.utoronto.ca}

\author[0000-0002-0278-7180]{Scott Tremaine}
\affiliation{Canadian Institute for Theoretical Astrophysics, Toronto, Ontario, Canada}
\affiliation{Institute for Advanced Study, Princeton, New Jersey, USA}

\date{\today}

\begin{abstract}
    Using the equations of motion from Hill's problem, with added accelerations for different forms of dynamical friction, we provide the (to-date) broadest scale-free study of friction-driven binary formation in gaseous disks and stellar clusters.
    We focus mainly on binary formation between stellar-mass black holes in active galactic nuclei (AGNi), considering both gas dynamical friction from AGN disks and stellar dynamical friction from the nuclear star cluster. We first find simple, dimensionless friction coefficients that approximate the effects of standard models for gas and stellar dynamical friction. We perform extensive simulations of Hill's problem under such friction, and we present a picture of binary formation through encounters between single stars on nearby orbits, as a function of friction parameter, eccentricity, and inclination. Notably, we find that the local binary formation rate is a linear function of the friction coefficient so long as the friction is weak. Due to the dimensionless nature of our model problem, our findings are generalizable to binary formation at all scales (e.g., intermediate-mass black holes in a star cluster, planetesimals in a gaseous disk).
\end{abstract}

\section{Introduction}

The census of observed black hole (BH) mergers continues to grow with successive observing runs of LIGO/Virgo. To better understand the various populations revealed by these observations, it is crucial to understand the origins of binary BHs (BBHs), as the formation process can establish parameters of a binary (orbital elements, spins, masses) that have observable signatures. Primordial BBHs --- those whose predecessor stars were in a  pair --- that have large enough separations to avoid overlap during the star-formation stage require longer than the age of the Universe to merge through gravitational radiation; thus studying the dynamical formation of BBHs is particularly important. (For a thorough review of this problem, see \citealp{Mandel_2022}.)

We provide an overview of the study of galactic nuclei as potential hosts to frequent BBH mergers (\S\ref{sec:state}) and discuss recent literature on the dynamical formation of BBHs in disks (\S\ref{sec:recent_lit}). In \S\ref{sec:this_work}, we detail the focus of the present work --- dynamical friction as a method of forming BBHs. Though we present this work as a study of BBHs in galactic nuclei, our model (\S\ref{sec:model}, \S\ref{sec:setup}) and results (\S\ref{sec:results}, \S\ref{sec:cap_rates}) are generalizable to binary formation in a broad variety of other environments. 

\subsection{Binary Mergers in Galactic Nuclei}\label{sec:state}
Galactic nuclei are the densest environments of stars and compact objects in a given galaxy \citep[e.g.,][]{Bahcall_1976, Merritt_2010, Merritt_2013, Hailey_2018, Gallego-Cano_2018}. Moreover, the density of BHs relative to ordinary stars is likely to be higher in galactic nuclei than in the rest of the galaxy, since massive BHs spiral towards the center of the galaxy due to stellar dynamical friction \citep[e.g.,][]{Miralda-Escude_2000}.

In their ``active'' states, these nuclei feature an accretion disk, at scales $\lesssim 0.1$--10 pc, surrounding a central (super)massive black hole ((S)MBH; e.g., \citealp{Lynden-Bell_1969,Pringle_1973,Soltan_1982}). In recent decades, these accretion disks in active galactic nuclei (AGNi) have drawn attention as possible nurseries of intermediate mass BHs \citep[e.g.,][]{McKernan_2012, McKernan_2014, Bellovary_2016} and BBH mergers \citep[e.g.,][]{Bartos_2017,Stone_2017,Secunda_2019,Yang_2019a,Grobner_2020,Ishibashi_2020,McKernan_2020,Secunda_2020,Tagawa_2020,SaavikFord_2022}. The argument motivating this attention requires several steps:

\begin{enumerate}
    \item In a given galaxy, both the absolute density of BHs and their density relative to ordinary stars are highest in the nucleus (see above).
    \item The orbits of BHs within a few pc of the SMBH may tend to become aligned with an AGN accretion disk through interactions with the gas \citep[e.g.,][]{Syer_1991,Rauch_1995,Macleod_2020,Fabj_2020}. The results of \citet{Fabj_2020}, \citet{Nasim_2023}, and \citet{Generozov_2023} suggest, for a wide range of initial BH orbits, that alignment with an AGN accretion disk could take longer than the disk's lifetime; however, all find that alignment is plausible for some non-negligible fraction of initial orbits. \citet{Wang_2024} present a thorough overview of alignment processes. Whatever alignment does occur further boosts the density of BHs within the accretion disk relative to the already-high density of the nucleus. 
    \item Within an accretion disk, torques from the surrounding, differentially rotating gas will act on orbiting BHs, similar to the torques on bodies in protoplanetary disks \citep[e.g.,][]{Goldreich_1980, Paardekooper_2010}. Migration “traps” (where inward- and outward-forcing torques from the gas disk balance out) can further enhance the density of BHs in localized regions of the disk (\citealp{Bellovary_2016, Secunda_2019, Yang_2019a, SaavikFord_2022}; see also \citealp{Grishin_2023}, who emphasize the importance of migration traps due to thermal torques).
    \item Given these arguments, we expect that some regions of AGN disks will host extremely high densities of \textit{single} BHs. In such regions, binary formation through dynamical processes --- which are always more efficient in regions with a higher density of single BHs --- could yield a substantial population of BBHs. Recent population studies have suggested that dynamical formation of BBHs is important in such regions \citep[e.g.,][and see \S\ref{sec:recent_lit}]{Tagawa_2020}, though the prescriptions for formation used in these studies deserve scrutiny \citep[e.g.,][]{Delaurentiis_2023}. 
    \item Any population of dynamically formed BBHs is likely supplemented by the remnants of a population of high-mass stellar binaries --- observational estimates suggest that the binary fraction of massive stars in the Milky Way nucleus is $\sim 30$\% \citep{Pfuhl_2014} --- though the evolution from binary stellar system to BBH is complicated and uncertain.
    \item When BBHs are present in AGN disks, they may be driven to inspiral by the von Zeipel-Kozai-Lidov mechanism \citep[e.g.][]{Hoang_2018, Fragione_2019}, by interactions with single BHs/stars \citep[e.g.,][]{Leigh_2018,Trani_2024, Fabj_2024}, and through energy dissipation into surrounding gas \citep[e.g.,][]{Escala_2005, Kim_2007, Baruteau_2011, Stone_2017, Tagawa_2018, Kaaz_2023, Dittmann_2024, Calcino_2023}. We note, however, that certain properties of the binary, the disk, and the local stellar population can lead binaries to \textit{out}spiral rather than inspiral (e.g., \citealp{Moody_2019, Tiede_2020, LiLai_2024, Trani_2024}; see also the recent review of circumbinary accretion by \citealp{Lai_2023}).
    \end{enumerate} 
Each of these steps comes with its own uncertainties; nevertheless, the arguments as a whole make a plausible case that AGNi are the sites of a significant fraction of BBH mergers. 

Conveniently, this hypothesis may be testable. First, BBH mergers in AGN disks could have distinct electromagnetic signatures accompanying their gravitational-wave (GW) emission (e.g., \citealp{Bartos_2017,Stone_2017,McKernan_2019,Kimura_2021,Tagawa_2023,Rodriguez-Ramirez_2023}), and there is one such candidate event --- GW 150914 --- among the first detected BBH mergers (\citealp{Graham_2020,Calderon-Bustillo_2021,Graham_2023,Morton_2023}; but see also \citealp{Palmese_2021,Ashton_2021}).
These outbursts may also efficiently produce neutrinos \citep{Tagawa_2023b,Zhu_2024}. Second, any set of processes leading to BBH mergers in AGNi would likely also lead to mergers of neutron star--neutron star and neutron star--BH binaries. The electromagnetic and neutrino-based signatures of such mergers in AGN disks are better understood \citep[e.g.,][]{Zhu_2021,Ren_2022}, and these could perhaps be used to constrain models of BBH formation and evolution in the same environment.

Merger detections through GW emissions alone can also serve to test this hypothesis. If a high fraction of all BBH mergers occur in AGNi, population-wide merger statistics would be skewed by the characteristics of these mergers (e.g., \citealp{Breivik_2016,Tagawa_2021, Samsing_2022, McKernan_2023, Vaccaro_2023, Trani_2024}). For example, BBH mergers with more unequal masses apparently tend to have more positive effective spins \citep[e.g.,][]{Callister_2021}, which several authors have attributed to the AGN merger pathway \citep[e.g.,][]{McKernan_2022,Santini_2023}. 
If the remnants of stellar-mass BBH mergers are not kicked significantly out of the plane of the disk, ``hierarchical'' mergers involving these more massive remnants may be more common within AGN disks than in other astrophysical environments \citep[e.g.,][]{Yang_2019a}; then prevalent GW detections of more massive BBH systems (e.g., those involving a BH or BHs within the pair instability mass gap), in concert with trends in mass ratio and effective spin, may suggest the AGN disk merger pathway is effective.
Finally, one can search for a spatial correlation between known AGNi and GW signals --- so far, \citet{Veronesi_2023} have used this tactic to suggest that most mergers do not come from the most luminous AGNi (though they make no claims about less luminous nuclei).

Although more and better observational data is essential, the simplest means of assessing the validity of this string of arguments is to better understand the theory behind each of them. We focus this work on one question at the core of item 4: how effective is the formation of binaries from single BHs within galactic nuclei?

\subsection{The Binary Formation Question}\label{sec:recent_lit}
The formation of a permanent binary in a hierarchical triple system (i.e., two nearby stellar BHs orbiting a more distant SMBH) presents a delicate dynamical problem. Energy must be removed near the pericenter of a hyperbolic encounter between two single BHs in order for them to become bound. 

Without dissipation, the two bodies might experience Jacobi capture --- a transient binary state with repeated close encounters \citep[e.g.,][]{Colombo_1966, Singer_1968, Petit_1986, Murison_1989, Goldreich_2002, Iwasaki_2007, Boekholt_2023}. The number of close encounters during a given interaction is a fractal function of the impact parameter of the two small bodies (see \citealp{Petit_1986}, or \citealp{Boekholt_2023} for high-resolution spectra of the number of close encounters across a range of impact parameters). 

If a pair of bodies has some means of dissipating energy during close passage, transient Jacobi captures could lead to permanent binary formation. For example, GW emission could provide the required energy dissipation if the closest approach distances are very small \citep[e.g.,][]{Hansen_1972, Li_2022, Boekholt_2023, Rom_2024}, though sufficiently close encounters are rare enough that this is unlikely to be the dominant binary formation mechanism \citep{Tagawa_2020}. A passing third body could remove the necessary energy \citep[as in several formation theories for Kuiper-belt binaries; see][]{Goldreich_2002, Schlichting_2008}, though the requirement of three bodies in close proximity means that this process would generally be less efficient than one involving only the two that will become a binary.

Within an AGN disk, a BH will experience a dissipative force from the surrounding gas \citep{Ostriker_1999} --- this is known as gas dynamical friction (GDF).  In a population synthesis study of compact object mergers in galactic nuclei, \citet{Tagawa_2020} used a friction timescale argument \citep[based on][]{Goldreich_2002} as a prescription for BBH formation in this environment. Essentially, for a given interaction between two single BHs, they compared the timescale of this gaseous dissipation (their $t_{\text{GDF}}$) to the time it would take the two BHs to pass through their mutual Hill radius (their $t_{\text{pass}}$). If the dissipation timescale was shorter, BBH formation was guaranteed; otherwise, the probability of BBH formation was set to $t_{\text{pass}}/t_{\text{GDF}}$ (see their eq.\ 64). This simple prescription, implemented in their very thorough modelling of dynamical processes in AGN disks, led to the conclusion that this gas-driven formation channel contributes up to 97\% of BBHs merging in this environment. 

Since this result, a bevy of studies have sought to better understand this formation process \citep{Li_2023, Rowan_2023, Delaurentiis_2023, Rozner_2023, Qian_2024,Whitehead_2023,Whitehead_2023b}.

\citet{Li_2023} and \citet{Rowan_2023} performed two- and three-dimensional global hydrodynamic simulations of AGN disks, respectively. Both showed that GDF-assisted capture is possible across a wide range of disk parameters and confirmed that gas friction can continue to harden BBHs post-capture \citep[cf.][]{LiLai_2022}. \citet{Whitehead_2023} used two-dimensional hydrodynamic, Hill's problem\footnote{Note that several of the papers referenced here use ``sheared sheet'' rather than ``Hill's problem'' to describe their setups --- for consistency, we use ``Hill's problem'' throughout.} simulations and studied a wider range of parameter space than \citet{Li_2023} or \citet{Rowan_2023}.

In these simulations, minidisks form around each single BH, and it is the dissipation of energy into these minidisks (and the circumbinary disk that forms when the two BHs strongly interact) that allows for binary formation. While the aforementioned studies each assumed the gas to be locally isothermal, \citet{Whitehead_2023b} allowed the \textit{adiabatic} mixture of gas and radiation in their simulations to respond to heating from this dissipation. Their results suggest that BBH formation is still efficient, despite the deposited energy puffing up the minidisks. (Interestingly, they also note that this deposition could create observable flares within the disk.)

\citet{Rozner_2023}, \citet{Delaurentiis_2023}, and \citet{Qian_2024} looked at capture with analytic prescriptions for GDF based on the linear analysis of \citet{Ostriker_1999}. \citet{Rozner_2023} found an analytic condition for capture in the isolated two-body case which depends on the relative velocities of the two bodies at large separations. \citet{Delaurentiis_2023} and \cite{Qian_2024} performed parameter-space studies that included the effect of a central SMBH and characterized the rate of binary formation for different sets of disk and BH parameters. The results of \citet{Delaurentiis_2023}, as the first to cover a substantial range of possible parameters, cast doubts on the simple formation prescription used in \citet{Tagawa_2020}.

All of these studies find that BBH formation via GDF should be common in AGN disks; however, there has been no clear, cohesive picture of the formation efficacy across the vast parameter space involved in this problem. Full-scale hydrodynamical simulations, which capture the detailed physics of interactions between the BHs and the surrounding gas, cannot provide full coverage of the parameter space. Coverage is more easily attained in simulations with analytical prescriptions for GDF, though these must make approximations that can limit their accuracy. We take the latter approach. In \S\ref{sec:compare}, we compare our results to those of recent semi-analytic and hydrodynamical works. This constitutes the first synthesis of these studies of binary formation through GDF.

We are also interested in other processes that act on BHs in galactic nuclei. Previous works have disregarded the possible contribution to BBH formation from \textit{stellar} dynamical friction (SDF). Compact nuclear star clusters (NSCs) appear at the centers of most galaxies \citep[for a review, see][]{NSC_review_2020}; our Galaxy has one \citep[for a review, see][]{Genzel_2010}, and they appear to be relatively common in all but the most massive galaxies \citep{boker_2009}. In galactic nuclei with an NSC around a SMBH, the total mass of the cluster stars is often comparable to the SMBH mass \citep[e.g.,][]{Hoyer_2024}. A high-mass NSC would exert  \citet{Chandrasekhar_1943} dynamical friction on a BH orbiting within it. In a similar manner to GDF, this force could dissipate energy during interactions between two single BHs, perhaps leading to the formation of a permanent binary. 

For now, we make two additional comments about SDF: (\textit{i}) Most galaxies host AGN accretion disks for only a small fraction of their lifetime ($\sim10^7$--$10^9$ yr; e.g., \citealp{Yu_2002,Marconi_2004}). If GDF drags BHs into the disk plane during an AGN phase (item 2 of the argument in \S\ref{sec:state}), then SDF can continue to act on the BHs and perhaps form binaries during the much longer period while the nucleus is inactive. (\textit{ii}) The stars in the NSC also contribute stochastic changes to the velocities of the BHs through close encounters. We have neglected these changes since they are generally smaller than the systematic accelerations due to GDF and SDF.

We discuss the relative importance of GDF and SDF in greater detail in \S\ref{sec:typical_frics}.

\subsection{This Work}\label{sec:this_work}

Here, we study the interactions of BHs in galactic nuclei using a simplified prescription for both GDF and SDF. We use Hill's problem as a framework for the dynamics (see, e.g., \citealt{Petit_1986} or \citealt{Tremaine_2023}), which describes the motion of two nearby bodies around a more distant and more massive one. (Formally, the masses $m$ and separation $r$ of the small bodies compare to the mass $M$ and distance $a$ of the large one as $m\ll M$, $r\ll a$, $m/r^3\sim M/a^3$.) The dimensionless nature of Hill's equations of motion allows us to generalize our results to a wide range of parameters (SMBH and BH masses, orbital radii, etc.). 

If we add dynamical friction to Hill's problem, we can still write dimensionless equations of motion (at least when the two small bodies have equal masses --- the case we focus on here) with one or more free parameters encoding the strength of the dynamical friction (which depends on the AGN disk density and sound speed, stellar density and velocity dispersion, etc.).  In the simplest case, the strength of GDF is set by a single coefficient $C_g$, which encodes all physical properties of the system,\footnote{Note that our coefficient $C_g$ is equivalent to $1/\tau_{\text{DF}}\Omega_K$ from \S6 of \citet{Delaurentiis_2023} and $1/\hat{\tau}\Omega_K$ of \citet{Qian_2024}, where $\tau$ is a dynamical friction timescale and $\Omega_K$ is the local Keplerian velocity. We discuss these works in more detail in later sections.} while the strength of SDF is set by a similar coefficient $C_s$. We derive these proportionalities in \S\ref{sec:model}. 

These prescriptions provide a simple framework in which we can study the effects of dynamical friction. We argue that the uncertainties in parametrizing the accurate, multi-parameter, non-linear effect of dynamical friction (as observed in hydrodynamic simulations; e.g., \citealp{Li_2023}, \citealp{Rowan_2023}, and \citealp{Whitehead_2023,Whitehead_2023b}) are large enough that such a simple prescription serves as a justifiable starting point for studying the efficacy of BBH formation. Furthermore, the many-order-of-magnitude uncertainties in our understanding of AGN disk structure (see, e.g., differences between models by \citealp{Sirko_2003}, \citealp{Thompson2005}, and \citealp{Hopkins_2024c}, with further discussion later) suggest that results based on these simple prescriptions would not introduce significant errors if used in population-synthesis models of BBHs forming within these disks.

Using this framework, we integrate more than $10^8$ interactions between equal-mass bodies on nearby, nearly circular and coplanar orbits, with impact parameters up to a few Hill radii and a range of eccentricities or inclinations, under a range of dynamical friction coefficients. We find that the binary formation rate depends roughly linearly on $C_g$ and $C_s$, at least down to friction coefficients $\sim 10^{-5}$ \citep[cf.][]{Qian_2024}. We also find that captures occur preferentially among impact parameters among and near those yielding Jacobi capture in the friction-free case \citep[see, e.g.,][]{Boekholt_2023}; this behavior has already been observed to some extent by \citet{Delaurentiis_2023} and \citet{Qian_2024}. Lastly, we present capture rates for encounters between BHs on eccentric and inclined orbits, which have not been studied in the literature on GDF so far. We aim to provide the (to-date) broadest scale-free study of binary formation assisted by dynamical friction in gaseous disks and stellar clusters.

In \S\ref{sec:model}, we derive the forms of our dynamical friction models; we detail the setup of our integrations and discuss interesting features and complications in \S\ref{sec:setup}. In \S\ref{sec:results} we provide examples of individual simulations and explore the dependence of capture on the friction coefficients $C_i$, the impact parameter of a given encounter, and the eccentricities $e$ and inclinations $I$ of the orbits of the single BHs. In \S\ref{sec:cap_rates}, we present formation rates as functions of $C_i$, $e$, and $I$, and we provide details of a dynamically formed population of binaries. Finally (\S\ref{sec:discussion}), we discuss our findings in the context of the many recent hydrodynamic and (semi-)analytic results mentioned above, and we briefly discuss realistic values of $C_i$ within galactic nuclei, presenting an overview of the current state of understanding of BBH formation under dynamical friction.

\section{Model}\label{sec:model}

We describe the equations of motion (EOM) we use to simulate interactions between single BHs under GDF and SDF.

\subsection{Hill's Problem, Unaltered}

Derivations of the EOM for Hill's problem are given by \cite{Petit_1986} and by \citet{Tremaine_2023}. Here, we provide a summary of the derivation.

Consider the motion of two bodies of mass $m_1$ and $m_2$ in nearly circular orbits around a central body of mass $M$ (which sits at the origin of our initial frame of reference). The bodies have position vectors $\mathbf{r}_1(t)$ and $\mathbf{r}_2(t)$; introduce a reference vector $\bar{\mathbf{a}}(t)$ of fixed magnitude from the origin to an arbitrary point near these bodies, and set $\bar{\mathbf{a}}$ to rotate uniformly with the local Keplerian velocity. We can write down the EOM of the vectors separating $\mathbf{r}_1$ and $\mathbf{r}_2$ from the reference vector, then simplify these equations by making Hill's approximations: (1) these separation vectors are much shorter than $|\bar{\mathbf{a}}|$, and (2) $m_1$, $m_2 \ll M$. We can then transform to a frame corotating with $\bar{\mathbf{a}}$, which has a constant angular speed $\Omega$ set by $M$ and $|\bar{\mathbf{a}}|$. In this frame, we can define $\Delta\mathbf{x}_1(t)$ and $\Delta\mathbf{x}_2(t)$ as the vectors separating the bodies from the reference vector (now effectively the origin), and we can write down EOM for these vectors. These are Hill's equations, which describe the motion of two nearby bodies, in distant orbits around a much larger body, relative to an arbitrary, nearby reference point. The only physical properties involved in these equations are $\Omega$ and $m_{1,2}$.

We can undertake another change of variables to find EOM for the barycenter of the two bodies $\mathbf{x}_{\text{cm}}$ and for their separation $\mathbf{x} \equiv \Delta\mathbf{x}_2 - \Delta\mathbf{x}_1$. Conveniently, the systems of equations for the barycenter and separation are separable; for our purposes, we are only interested in the separation, so we can ignore the EOM of the barycenter. As we describe below, if dynamical friction is included the EOM remain separable, but only for equal-mass bodies, $m_1=m_2$.

Finally, we can make this problem dimensionless by transforming from $t \rightarrow t_d$, where $t_d \equiv \Omega t$, and from $\mathbf{x}\rightarrow \pmb{\rho}$, where $\pmb{\rho} = (\xi, \eta, \zeta) \equiv (x,y,z)/3^{1/3}R_{\text{H}}$, with the mutual Hill radius 
\begin{equation}
    R_{\text{H}} = \bar{a}\left(\frac{m_1+m_2}{3m_0}\right)^{1/3}.
\label{eq:hilldef}
\end{equation}
This procedure yields the dimensionless EOM for the vector separating the two bodies:
\begin{equation}\label{eq:hill_eom_dimless}
    \pmb{\rho}'' = 2\pmb{\rho}''_{\text{cor}} + \pmb{\rho}''_{\text{cen}} - \frac{\pmb{\rho}}{\rho^3},
\end{equation}
defining $\pmb{\rho}''_{\text{cor}} \equiv (\eta',\,-\xi',\,0)$ and $\pmb{\rho}''_{\text{cen}} \equiv (3\xi,\,0,\,-\zeta)$ (the subscripts ``cor'' and ``cen'' stand for ``Coriolis'' and ``centrifugal'' accelerations), and denoting $\rho = |\pmb{\rho}|$ as the distance between the two bodies in these dimensionless coordinates. Primes denote derivatives with respect to $t_d$.

These are the unaltered EOM for the dimensionless Hill's problem. We now seek to add the effect of GDF and SDF to the dynamics described by these equations.

\subsection{Gas Dynamical Friction}\label{sec:gdf}

According to \citet{Ostriker_1999}, the force of dynamical friction on a perturber of mass $m$ moving linearly through a gaseous medium is 
\begin{equation}\label{eq:GDF_base}
    \mathbf{F}_{\text{GDF}} = -\frac{4\pi(\mathbb{G}m)^2\rho_g}{v_{\text{rel}}^3}I\left(\mathcal{M}\right)\dot{\mathbf{x}}_{\text{rel}},
\end{equation}
where $v_{\text{rel}} = |\dot{\mathbf{x}}_{\text{rel}}|$ is the magnitude of the velocity vector of $m$ relative to the gas, $\rho_g$ is the local unperturbed mass density of the medium, and $I$ is a function depending on the Mach number $\mathcal{M} \equiv v_{\text{rel}} / c_s$, with $c_s$ the local sound speed. In the subsonic regime, $I$ is given by equation (14) of \citet{Ostriker_1999}; a Taylor expansion at $\mathcal{M} = 0$ yields $I = \mathcal{M}^3/3 + \mathcal{O}(\mathcal{M}^4)$. Plugging this expansion into equation (\ref{eq:GDF_base}), we find 
\begin{equation}
    \mathbf{F}_{\text{GDF}} = -k\dot{\mathbf{x}}_{\text{rel}},\quad \text{with} \quad k = \frac{4\pi (\mathbb{G}m)^2 \rho_g}{3c_s^3}.
\end{equation}

In terms of the coordinates of Hill's problem, we can say that the acceleration from GDF on the $i$th body of interest is $\Delta\ddot{\mathbf{x}}_{\text{GDF},\,i} = -(k/m_i) \left(\Delta\dot{\mathbf{x}}_i - \Delta\dot{\mathbf{x}}_{K,\,i}\right)$, where $\Delta\dot{\mathbf{x}}_{K,\,i} \equiv -(3/2)\Omega x_i\hat{\mathbf{y}}$ is the local Keplerian velocity and $i=1,2$. The acceleration of the separation is given simply by $\ddot{\mathbf{x}}_{\text{GDF}} = \Delta\ddot{\mathbf{x}}_{\text{GDF},\,2} - \Delta\ddot{\mathbf{x}}_{\text{GDF},\,1}$; however, the right-hand side of this cannot be written in terms of \textit{only} the separation vector (i.e., there is a term proportional to $\dot{\mathbf{x}}_{\text{cm}}$) unless we set $m_1 = m_2 \equiv m$, as we shall do henceforth. In the equal-mass case, $\ddot{\mathbf{x}} = -(k/m)\left(\dot{\mathbf{x}} - \dot{\mathbf{x}}_K\right)$. Converting this to dimensionless coordinates $(\pmb{\rho},t_d)$ yields
\begin{equation}\label{eq:GDF_final}
    \pmb{\rho}''_{\text{GDF}} = -C_g\left(\pmb{\rho}' + \frac{3}{2}\xi\hat{\pmb{\eta}}\right), \quad \text{with} \quad C_g = \frac{k}{\Omega m}.
\end{equation}
We add this term to the right-hand side of equation (\ref{eq:hill_eom_dimless}) to find the EOM for interactions that include GDF. 

The expression for the friction coefficient $C_g$ can be rewritten in an approximate form that provides more physical insight. We can set the sound speed $c_s$ by approximating the equation for vertical hydrostatic equilibrium as $p_g \approx \rho_g h^2\Omega^2$, with $p_g$ the pressure and $h$ the scale height, then assuming that the gas is isothermal so $c_s = \sqrt{p_g/\rho_g}$. These approximations yield $c_s \approx h\Omega = (h/r) v_K$, with $v_K$ the local Keplerian velocity. Plugging in expressions for $\Omega$ and $v_K$, one finds the dimensionless expression 
\begin{equation}\label{eq:K}
    C_g \simeq \frac{4\pi}{3}\frac{\rho_g m r^3}{(h/r)^3 M_{\bullet}^2},
\end{equation}
where $M_{\bullet}$ is the mass of the central body (in the case of an AGN accretion disk, the SMBH).

We repeat the principal approximations that we have used to derive this formula. (1) We have assumed that the relative velocity $v_\mathrm{rel}$ of the two small bodies is much less than the sound speed $c_s$. Using this approximation underestimates the GDF force at $\mathcal{M} \sim 1$, and overestimates it for $\mathcal{M} \gg 1$, but typically only by a factor of order unity over the range of $\mathcal{M}$ relevant here. (2a) The \citet{Ostriker_1999} prescription was derived in the context of an isolated perturbing body, so it may give misleading results once the BHs are strongly interacting (when their separation $x$ satisfies $\mathbb{G}m/x > c_s^2,v_\mathrm{rel}^2$). A fuller understanding of friction in this regime requires hydrodynamical simulations. (2b) This prescription for GDF neglects the net force on a BH due to accreting gas.\footnote{ We note the recent work of \citet{Suzuguchi_2024}; among other results, they suggest that 
the strength of GDF in simulations that include accretion is within a factor of a few of the \citet{Ostriker_1999} prescription (see their Figure 6). \citet{Rowan_2023,Rowan_2024} similarly showed that accretion makes little difference in the process of binary formation from GDF.
} 
(3) We have assumed that the two BHs have the same mass, such that the EOM for the center of mass and the relative separation are decoupled. In the case of BHs of unequal mass there will be an additional EOM describing the motion of the center of mass.

Despite these approximations, we believe that we have posed a simple problem that captures much of the physics of the formation of BBHs in accretion disks through GDF. 

\subsection{Stellar Dynamical Friction}\label{sec:sdf}

From equations (7.92) and (8.6) of \citet{Binney_2008}, the dynamical friction force acting on a body of mass $m$ moving with velocity vector $\dot{\mathbf{r}}$ through a field of bodies with mass $m_a$ and a Maxwellian velocity distribution with zero mean can be written
\begin{equation}\label{eq:SDF_base}
    \mathbf{F}_{\text{SDF}} = -\frac{4\pi\mathbb{G}^2\rho_s m(m+m_a)\ln\Lambda}{\sigma^2}G(X)\frac{\dot{\mathbf{r}}}{v},
\end{equation}
where $G$ is a function of $X\equiv v/\sqrt{2}\sigma$, which is itself a function of the magnitude of the velocity vector of the body of interest $v = |\dot{\mathbf{r}}|$ and the velocity dispersion of the field stars $\sigma$; $\rho_s$ is the local field density; and $\ln\Lambda$ is the Coulomb logarithm, with $\Lambda$ defined in equation (7.84) of \citet{Binney_2008}.

As in the GDF derivation, we work with equal-mass bodies.\footnote{In contrast to GDF, bodies of unequal mass on circular orbits subjected to SDF will spiral toward the central black hole at different rates. This differential inspiral should not strongly affect the binary formation rate.} Taking the differences between equation (\ref{eq:SDF_base}) for the two bodies, we can write
\begin{equation} \label{eq:SDF_diff}
    \ddot{\mathbf{r}}_2 - \ddot{\mathbf{r}}_1 \propto \frac{G(X_2)}{v_2}\dot{\mathbf{r}}_2 - \frac{G(X_1)}{v_1}\dot{\mathbf{r}}_1,
\end{equation}
with a factor in front encoding all physical values unrelated to the velocities of the bodies of interest. We define $\dot{\mathbf{r}}_{1,2} = \dot{\mathbf{r}}_{\text{cm}} \pm \Delta\dot{\mathbf{r}}$, where $\mathbf{r}_{\text{cm}}$ denotes the vector from the central body to the barycenter of the objects of interest. We assume that all non-Keplerian components of the velocity are negligible and that the difference in the bodies' Keplerian velocities is small ($\mathbf{v}_{K2} \simeq \mathbf{v}_{K1} \equiv \mathbf{v}_{K}$). Expanding equation (\ref{eq:SDF_diff}) to first order around $\Delta\dot{\mathbf{r}} = 0$, these approximations yield 
\begin{multline}
    \ddot{\mathbf{r}}_2 - \ddot{\mathbf{r}}_1 \propto \frac{1}{v_K}\Big[\Gamma_1(\dot{\mathbf{r}}_2 - \dot{\mathbf{r}}_1) \\ + (\Gamma_2 - \Gamma_1)(v_{2y} - v_{1y})\hat{\mathbf{e}}_y\Big],
\end{multline}
where $\Gamma_1$ and $\Gamma_2$ are dimensionless constants dependent on the distribution and velocity dispersion of the field bodies. We can make this differential acceleration dimensionless to find an equation of the form 
\begin{equation}\label{eq:SDF_final}
    \pmb{\rho}''_{\text{SDF}} = - C_s\,\Big[\,\Gamma_1\xi'\hat{\pmb{\xi}} 
    + \Gamma_2\eta'\hat{\pmb{\eta}}
    + \Gamma_1\zeta'\hat{\pmb{\zeta}}\,\Big],
\end{equation}
with $C_s$ a constant dependent on the properties of the sea of bodies.

Here, we assume the stellar distribution in the NSC is spherically symmetric and ergodic, following a mass density profile $\rho_s(r) \propto r^{-\gamma}$. With this, one finds
\begin{align} \label{eq:C}
    C_s &= \frac{4\pi r^3\rho_s(m+m_a)\ln\Lambda}{M_{\bullet}^2}
    \\
    \Gamma_1 &= \text{erf}\left(\sqrt{\frac{\gamma+1}{2}}\right)
    \nonumber \\
    &\quad\quad - \sqrt{\frac{2(\gamma + 1)}{\pi}}\exp\left(-\frac{\gamma+1}{2}\right)
    \\ 
    \Gamma_2 &= -2\,\text{erf}\left(\sqrt{\frac{\gamma+1}{2}}\right) + \Bigg(2\sqrt{\frac{2(\gamma + 1)}{\pi}} 
    \nonumber \\ \label{eq:gamma_2}
    &\quad \quad + \sqrt{\frac{2(\gamma + 1)^3}{\pi}}\Bigg)\exp\left(-\frac{\gamma+1}{2}\right).
\end{align}
We expect that $\gamma$ will be between 1.5 and 2.5 \citep[e.g., profiles in][]{Pechetti_2020}. In this range, $\Gamma_1$ varies from $0.21$ to $0.19$ and $\Gamma_2$ varies from $-0.06$ to $-0.13$. For our simulations below, we take $\gamma = 1.5$.

Appendix \ref{sec:appendix} includes a full derivation of equation (\ref{eq:SDF_final})  and the coefficients $\Gamma_1$ and $\Gamma_2$. As in the case of GDF, this prescription hinges on a number of assumptions, but serves as a reasonable starting point for understanding the dissipative effects of a spherically distributed, collisionless sea of bodies during binary interactions. 

\subsection{A Comment on these Friction Equations}

The expressions for GDF (eq.\ \ref{eq:GDF_final}) and SDF (eq.\ \ref{eq:SDF_final}) are, at face value, quite similar in form. They both include a coefficient describing the overall ``strength'' of dynamical friction ($C_g$ for gas and $C_s$ for stars), and in both cases the acceleration is linearly proportional to velocity. That said, the differences in how these frictional forces act on the azimuthal velocity $\eta'$ can lead to substantially different dynamics. GDF only acts to damp any velocity relative to the local Keplerian velocity (i.e., the GDF vanishes if $\pmb{\rho}'=-(3/2)\xi\hat{\pmb{\eta}}$). In contrast, SDF acts even on objects on circular orbits. SDF decreases $|\eta'|$, which in turn produces an inequality between the $\hat{\pmb{\xi}}$ components of $\pmb{\rho}''_{\text{cor}}$ and $\pmb{\rho}''_{\text{cen}}$ (see eq.\ \ref{eq:hill_eom_dimless}) that shrinks the radial separation between two bodies on nearby, initially circular orbits. If the friction is strong enough (see \S\ref{sec:preint_effects}), a significant radial drift arises long before the bodies begin to interact with each other. 

In the sections below, we conduct our integrations and present our results with arbitrary values of $C_g$ and $C_s$ so that they are as general as possible (with the exception of our choice of $\gamma$). We discuss realistic values of these parameters in galactic nuclei in \S\ref{sec:typical_frics}.

\section{Integrations}\label{sec:setup}

To study each type of dynamical friction, we integrate Hill's dimensionless EOM (eq.\ \ref{eq:hill_eom_dimless}) with the relevant acceleration term (eq.\ \ref{eq:GDF_final} or \ref{eq:SDF_final}) added to the RHS. This system of equations must be solved numerically. We use the DOP853 algorithm \citep{DOP853} as implemented in \href{https://docs.scipy.org/doc/scipy/reference/generated/scipy.integrate.DOP853.html}{\texttt{scipy.ode}}. DOP853 is an explicit Runge-Kutta solver of order 8(5,3); we set the relative and absolute error tolerances to \texttt{rtol} $=$ \texttt{atol} $= 10^{-10}$. Other integrators (e.g., LSODA, dopri5) returned consistent results.

\subsection{Initial Conditions}\label{sec:ICs}

We integrate two general classes of interaction: (1) those between two bodies on circular orbits in a common plane, and (2) those between bodies on eccentric and/or inclined orbits.

For class (1), we set the initial radial separation $\xi_0 \equiv f_\xi r_{\text{H}}$ and the initial azimuthal separation $\eta_0 \equiv f_\eta r_{\text{H}}$, where $r_{\text{H}} = 3^{-1/3}$ is the mutual Hill radius in the dimensionless Hill's problem. The factors $f_\xi$ and $f_\eta$ can be chosen to be positive; we randomly draw the impact parameter $f_\xi$ for each interaction from a uniform distribution over $[0.5,3.0]$, though we vary this domain slightly under certain conditions in order to probe the full range of initial conditions (ICs) leading to capture (e.g., for SDF with $|C_s\Gamma_2| > 10^{-2}$ and for highly-eccentric tests, we expand the window to higher $f_\xi$). We choose $f_\eta = 20$ for all interactions --- this choice is somewhat arbitrary (cf.\ \S\ref{sec:preint_effects}), though we would like $\eta_0$ to be large enough that the mutual perturbations between the two small bodies are negligible at the start of the integration. Note that these ICs are less accurate than the asymptotic approximation given in \citet{Petit_1986}; we assume that the overall capture cross-section will be robust to slight adjustments to the ICs of individual interactions. We complete our set of ICs for circular, flat orbits by setting $\xi'_0 = \zeta = \zeta' = 0$ and $\eta'_0 = -\frac{3}{2}\xi_0$.

For class (2), we consider bodies on orbits that are eccentric or inclined relative to each other. First, imagine $m_2$ is initially on an eccentric/inclined orbit. If we assume that the two bodies are not initially interacting, $m_2$ will undergo epicyclic motion in the frame of $m_1$. We can write down the exact solution for this motion relative to the reference vector $\bar{\mathbf{a}}$, given as $\Delta\mathbf{x}_2(t)$ \citep[e.g.,][equation 3.104]{Tremaine_2023}. Defining the initial motion of $m_1$ in the same way, we can find the separation vector $\mathbf{x}(t)$ and convert it to dimensionless coordinates. Setting $t=0$ at the start of the integration, we obtain the set of ICs
\begin{align}
    \pmb{\rho}_0
    &=
    \begin{pmatrix}
        f_\xi - f_e\cos\delta_e
        \\
        f_\eta + 2f_e\sin\delta_e
        \\
        f_I\cos\delta_I
    \end{pmatrix} r_{\text{H}}
\end{align}
and
\begin{align}
\quad \pmb{\rho}'_0
    &=
    \begin{pmatrix}
        f_e\sin\delta_e
        \\
        -\frac{3}{2}f_\xi + 2f_e\cos\delta_e
        \\
        -f_I\sin\delta_I
    \end{pmatrix} r_{\text{H}}.
\end{align}
That is, the separation vector is undergoing epicyclic motion that is separable into a radial/azimuthal component (i.e., in the $\xi$--$\eta$ plane) and a vertical component, with a radial/azimuthal amplitude $f_e$ and a vertical amplitude $f_I$; we start a simulation at phases of these epicycles given by $\delta_e$ and $\delta_I$.

We test a range of values of $f_e$ and $f_I$, which we define as $f_e \equiv e\left(\bar{a}/R_{\text{H}}\right)$ and $f_I \equiv I\left(\bar{a}/R_{\text{H}}\right)$; however, we stress that the parameters $e$ and $I$ are not always intuitive. In the case where one body has non-zero eccentricity or inclination (while the other is on a circular, flat orbit), $e$ is the eccentricity and $I$ the inclination of that body. When \textit{both} bodies have non-zero eccentricity (inclination), $e$ ($I$) is not a simple function of the eccentricities (inclinations), but is set by the difference of the Runge-Lenz vectors of the individual BHs. We present BBH formation rates for interactions for interactions with arbitrary $f_e$ and $f_I$, but some care is needed to apply these results to populations of single BHs in which both members of an interacting pair are on eccentric, inclined orbits (see also \S\ref{sec:calculate}). 

We expect bodies on nearby orbits to encounter each other at random phases, so we draw $\delta_e$ and $\delta_I$ from a uniform distribution over $[0,2\pi]$. We choose $f_\xi$ and $f_\eta$ in the same manner as for class (1). 

\subsection{Pre-interaction Dissipation}\label{sec:preint_effects}

In simulating interactions with dynamical friction, energy is dissipated in the leadup to the first strong interaction between the bodies, so our ``arbitrary'' choice of the initial azimuthal position $f_\eta$ could have a significant impact on our results. We here consider this effect for our two classes (1, circular and flat orbits, and 2, eccentric or inclined orbits) and for both GDF and SDF. 

We define $t_1$ as the (dimensionless) time it takes for the bodies to reach $\rho = r_H$ in a given integration. This is well-approximated by $t_1 \simeq 2\eta_0/3\xi_0$ (i.e., the time to travel a distance $\eta_0$ at the Keplerian velocity, holding $\xi(t) = \xi_0$). We also define a (dimensionless) timescale for frictional damping in the absence of interactions, $\tau_f \equiv 1/C_g$ for GDF \citep[cf.][]{Delaurentiis_2023,Qian_2024} and $\tau_f \equiv 1/|C_s\Gamma_2|$ for SDF (see below for more on this correction). Finally, we define $t_{f}$ as the (dimensionless) time over which friction can act on the system prior to the onset of strong interactions, which is not necessarily the same as $t_1$ --- again, see below. Roughly speaking, when $\tau_f \gg t_f$, the pre-interaction effects of dynamical friction should be negligible, and the choice of the initial azimuth $f_\eta$ $(\propto \eta_0)$ should not affect our results (so long as it is large enough; e.g., \citealp{Petit_1986}). 

When GDF is present, there is no friction so long as the approaching bodies are on circular orbits in the midplane of the gas disk. Thus, for class (1) interactions, GDF acts only after non-circular components of the velocity are introduced by the mutual gravitational interaction of the two bodies. From numerical tests, we estimate that such velocities become substantial when the system reaches $\eta \sim 5r_H$ (e.g., Figure 10 of \citealp{Qian_2024}), so friction can act over a time $t_f \sim 2(5r_H)/3\xi_0$. We always begin our numerical integrations at $\eta_0=20r_H$, so $t_1=40 r_H/3\xi_0$ and $t_f < t_1$ always; thus the pre-interaction dissipation should be independent of our exact choice of $f_\eta$. When $\tau_f \sim t_f$ --- i.e., when the dissipation timescale is of order the time during which GDF will act on class (1) cases prior to the onset of strong gravitational interactions --- we anticipate substantial pre-interaction dissipation. We expect that this dissipation will be negligible when $C_g \ll 1/t_f \sim 0.3\,(\xi_0/r_H)$; for $C_g$ greater than this, GDF likely suppresses the non-Keplerian motion that would arise due to gravitational interactions between the two bodies as they approach $\rho = r_H$.

In contrast, with SDF $t_f = t_1$ because SDF is present even if the bodies are traveling on circular orbits --- the terms proportional to $\Gamma_1$ in equation (\ref{eq:SDF_final}) cancel, but the term proportional to $\Gamma_2$ does not. The strength of SDF in this situation is then effectively $|C_s \Gamma_2|$. (Note that $\Gamma_2 = -0.058$ for our fiducial choice of $\gamma = 1.5$.) At large separations ($\rho \gg r_H$), for Keplerian ICs with positive $\xi$ and $\eta$, SDF will act first to decrease $\eta'$ (the azimuthal velocity) on a timescale $\tau_f \approx 1/|C_s\Gamma_2|$; this will introduce an acceleration in the $-\xi$ direction (radially inwards). In other words, SDF causes differential radial drift of the two bodies. Thus when the SDF timescale is short, the approximation that the two bodies are initially on nearby orbits of constant radius is not a good one, and our results are likely to depend on our arbitrary choice of $\eta_0 = 20r_H$. In practice we believe that this drift should not compromise our results so long as $\tau_f \gg t_f = t_1$, which corresponds to $|C_s\Gamma_2| \ll 0.075\,(\xi_0/r_H)$.

In class (2) cases, the non-circular motions of well-separated bodies decay exponentially with a time constant of roughly $C_g^{-1}$ or $|C_s\Gamma_2|^{-1}$. For example, consider the initial motion of a system subject to GDF, with inclination amplitude $f_I \neq 0$. While $\rho \gg r_H$, the $\zeta$-component of the separation will evolve according to $\zeta'' + C_g\zeta' + \zeta = 0$, which yields a solution in which the inclination decays with time as $\exp(-C_gt_d/2)$. The analogous eccentric case yields a set of coupled differential equations in $\xi$ and $\eta$, and the solution of these equations shows that eccentricity decays as $\exp(-C_gt_d)$. Thus the inclinations and eccentricities of all the single BHs orbiting in the disk decay between encounters, and other processes such as gravitational scattering must excite them to maintain a dynamical equilibrium in which the orbits are non-circular. Exploring the nature of this equilibrium is beyond the scope of this paper.

\subsection{Simulations}

For our initial suite of simulations of class (1) interactions, we examine the range $C_g\in(10^{-5},10^0]$ and $C_s \in [10^{-4},10^{1}]$ (with $\gamma = 1.5$, so $\Gamma_2 = -0.058$) at logarithmically equal increments, as well as $C_i = 0$ as a control. We integrate the EOM $n = 10^5$ times for each value of the coefficient $C_i$; each integration begins with a randomly drawn impact parameter $f_\xi \equiv \xi_0/r_H \in [0.5,3.0]$. 

For non-circular (class 2) interactions, we only test $C_g = [10^{-3},10^{-2.5},10^{-2},10^{-1.5},10^{-1}$].\footnote{As we will see below, the capture rates under GDF and SDF are quite similar in regimes of $|C_s\Gamma_2|$ for which we trust our SDF prescription, even though the dynamics involved in pre-interaction dissipation under the two forms of friction differ substantially. Due to the expansive parameter space involved, we have restricted our class (2) analyses to GDF; we suggest caution in extrapolating these results to SDF.} We initially test a range of eccentricity and inclination amplitudes $f_e, f_I \in [0.0,1.5]$ at linearly equal increments. At each value of $f_e$ and $f_I$, we integrate the adjusted EOM $n = 10^2/C_g$ times with a randomly drawn impact parameter $f_\xi$ --- each of these interactions also has its own randomly drawn phases $\delta_e$ and $\delta_I$.

After initial test runs, we proceed to integrate more ICs at the same test values of $C_i$, $f_e$, and $f_I$, with the aim of attaining at least $5$ captures for each value tested. We also test higher values of $f_e$ and $f_I$. We eventually reach a floor in $C_i$ (and ceilings in $f_e$ and $f_I$) below (above) which the goal of 5 captures is too computationally expensive. All told, for class (1), we integrate $36.7$ million interactions under GDF and $24.2$ million interactions under SDF. For class (2), which are all under GDF, we integrate $52.2$ million eccentric interactions and $13.0$ million inclined interactions.

During each integration, at steps of $\Delta t_d = 0.1$, we track the Jacobi--Hill constant
\begin{equation} \label{eq:E_H}
    E_{\text{H}} = \frac{\pmb{\rho}'\cdot\pmb{\rho}'}{2} - \frac{1}{\rho} - \frac{3}{2}\xi^2 + \frac{1}{2}\zeta^2,
\end{equation}
which is conserved in the absence of dynamical friction; the separation $\rho = |\pmb{\rho}|$; and the binary semi-major axis, which is given in dimensionless coordinates by
\begin{equation}
    a_b = -\frac{1}{2}\left(\frac{\pmb{\rho}'\cdot\pmb{\rho}'}{2} - \frac{1}{\rho}\right)^{-1}.
    \label{eq:ab}
\end{equation}
Note that $a_b$ is generally positive for bound orbits and negative for unbound orbits.

We are also interested in the orbital properties of the BBHs formed by this process. For cases meeting the formation criteria defined below, we record the eccentricity $e_b \equiv [1 - l_b^2/\mathbb{G}(m_1+m_2)a]^{1/2}$, where $l_b = |\mathbf{l}_b| = \Omega_K^2(3^{1/3}R_H)^4|\pmb{\rho}\times\pmb{\rho}'|^2$ is the orbital angular momentum of the binary and $a = (3^{1/3}R_H)a_b$ is the semimajor axis \textit{with} dimensionality. We also record the inclination $\theta_b = \arccos(l_{b\zeta} / |\mathbf{l}_b|)$, which we define as the misalignment between the binary angular-momentum vector and the $\hat{\pmb{\zeta}}$ unit vector normal to the midplane of the disk. Note that this will always be $0$ or $\pi$ (i.e., perfectly prograde or retrograde) for cases with $f_I = 0$.

\subsection{Formation Conditions}\label{sec:capture}

The Jacobi--Hill constant is the Hill's problem analog to the binary energy --- as the name suggests, it is constant under Hill's EOM in the absence of dynamical friction. If dynamical friction is always dissipative, in the sense that $E_{\text{H}}' < 0$ for all combinations of $\pmb{\rho}$ and $\pmb{\rho}'$, we can be confident that a permanent binary has formed if $E_{\text{H}} < E_{\text{H},\,\text{L}} \equiv -3^{4/3}/2$ --- the value of the Jacobi--Hill constant of a stationary body at the Lagrange points $(\xi,\eta,\zeta) = (\pm r_{\text{H}},0,0)$. 

For our prescription for SDF, we can show that
\begin{equation}   E'_{\text{H,}\,\text{SDF}} = - C_s\left(\Gamma_1\xi'^2 + \Gamma_2\eta'^2 + \Gamma_1\zeta'^2\right),
\end{equation}
which 
can be positive when $\eta'$ dominates the velocity vector, as $\Gamma_2$ is negative for our fiducial NSC density slope parameter $\gamma = 1.5$.
Similarly, with our prescription for GDF,
\begin{equation}\label{eq:EH_dtd}
    E'_{\text{H,}\,\text{GDF}} = -C_g \left(\pmb{\rho}'\cdot\pmb{\rho}' + \frac{3}{2}\xi\eta'\right),
\end{equation}
which is only negative when $\rho'^2 > -(3/2)\xi\eta'$.\footnote{Hydrodynamical simulations show that GDF can add energy to a binary system (e.g., \citealt{Li_2023}). While our approximate prescription is likely not capturing the physics involved in that process, it is an encouraging feature that our prescription for GDF can add energy as well.} We therefore need a stronger condition for permanent capture than simply $E_{\text{H}} < E_{\text{H},\,\text{L}}$.

If the binary semi-major axis $a_b$ is small enough, the binary will be stable to small increases in $E_{\text{H}}$ (as argued in, e.g., \citealp{Boekholt_2023,Delaurentiis_2023,Qian_2024}). Thus we say that a binary capture has occurred if $a_b$ falls below $f_cr_{\text{H}}$ (with $f_c\lesssim 0.1$) while $E_{\text{H}} < E_{H,L}$. We have found that our results are robust to small changes in $f_c$. In this paper we use $f_c = 0.05$. 

On the other hand, we consider an interaction to be an ``escape'' when the separation $\rho$ exceeds some limit $f_r r_{\text{H}}$, indicating that the bodies have concluded their interaction and moved away from each other. We use the conservative value $f_r = 25$.

Note that our formation conditions consider the semimajor axis $a_b$ in the frame of the binary, while, for example, \citet{Qian_2024} use a condition based on the semimajor axis in the \textit{inertial} frame. These two definitions of $a_b$ are nearly the same at sufficiently small values.

\subsection{Calculating Formation Rates}
\label{sec:calculate}
Each integration ends when the system meets the condition for either capture or escape. We would like to calculate a formation rate, for a particular friction coefficient $C_g$ or $C_s$, based on these Boolean results for an ensemble of integrations.

In a disk of single BHs, we can define $N$ as the surface number density of BHs that will encounter each other with a characteristic $f_e$ and $f_I$. Then the rate of binary formation per unit area is given by \citep{Qian_2024}
\begin{equation}
    \mathcal{R}(C_i,f_e,f_I) = N\int_0^\infty F_{\text{cap}}(b\,;\,C_i,f_e,f_I)\cdot Nv_sdb.
    \label{eq:caprate}
\end{equation}
We have here introduced $b$, the (physical units) difference between the guiding center radii of the two initial orbits; in the dimensionless Hill's problem, this scales to $f_\xi$. Then $v_s =(3/2)\Omega_K b$ is the shear velocity and $Nv_sdb$ is the rate at which a single object encounters others with guiding center radius differences in $[b,\,b+db]$. The dimensionless number $F_{\text{cap}}(b)$ is the fraction of encounters in that range, with a given $f_e$ and $f_I$, leading to capture under a given friction coefficient $C_i$.

The local formation rate per unit area for a given $C_i$ is found by marginalizing $\mathcal{R}(C_i,f_e,f_I)$ over the local distribution of $f_e$ and $f_I$.\footnote{We note that the local distribution of $f_e$ and $f_I$ is not the same as the distribution of eccentricities and inclinations of single BHs in the disk. The distribution of $f_e$ and $f_I$ is given by the distribution of the \emph{differences} in the eccentricity and inclination vectors of pairs of bodies drawn randomly from the disk.} The total formation rate in a disk is found by integrating the formation rate per unit area in (\ref{eq:caprate}) over the area of the disk, considering the value of $C_i$ and the corresponding $\mathcal{R}$ as a function of position. The properties of AGN disks are not well-constrained (see discussion in \S\ref{sec:typical_vals}), so in this paper we will focus on the local formation rates as a function of friction coefficients

In the sections below, we report estimates of the (dimensionless) scaled capture rate
\begin{align}\label{eq:rate_scale}
    \rate(C_i,f_e,f_I) &\equiv \frac{\mathcal{R}(C_i,f_e,f_I)}{N^2\Omega_KR_H^2} 
    \nonumber \\
    &= \frac{3}{2}\int_{\xi_{0,\text{min}}/r_H}^{\xi_{0,\text{max}}/r_H}  F_{\text{cap}}(\xi_0)\cdot \frac{\xi_0 d\xi_0}{r_H^2},
\end{align}
where $\xi_{0,\text{max}}$ and $\xi_{0,\text{min}}$ are the upper and lower bounds, respectively, of our sampled range of impact parameters. This is equivalent to $\mathcal{R}/n^2\Omega_KR_H^2$ in the notation of \citet{Qian_2024}. Note that the bounds of this integral should in fact be $\pm \infty$ --- in setting stricter bounds, we have assumed that there are no captures outside our sampled range.

Our simulations provide a Monte Carlo integration of the integral in equation (\ref{eq:rate_scale}). For the $j^\mathrm{th}$ out of $n$ EOM integrations with a given friction coefficient $C_i$, which has impact parameter $\xi_{0,j}$ and a characteristic $f_e$ and $f_I$, we define $B_j \equiv 1$ if a BBH is formed and $B_j \equiv 0$ if not. Then we can estimate the scaled capture rate as
\begin{equation}\label{eq:rate_approx}
    \rate(C_i,f_e,f_I) \simeq \frac{3}{2} \frac{\Delta\xi_0}{r_H^2} \frac{1}{n}\sum_{j=0}^n B_j\xi_{0,j},
\end{equation}
where $\Delta \xi_0 = \xi_{0,\text{max}} - \xi_{0,\text{min}}$.

At every coefficient, we calculate the standard deviation of our rate estimates $\sigma_{\rate}$ by the typical formula for Monte Carlo integration; that is, 
\begin{equation}\label{eq:rate_error}
    \sigma^2_{\rate} \simeq \frac{9\Delta \xi_0^2}{4r_H^4}\frac{\langle f^2 \rangle - \langle f \rangle^2}{n-1},
\end{equation}
with $\langle f \rangle \simeq n^{-1}\sum_j B_j\xi_{0,j}$ and $\langle f^2 \rangle \simeq n^{-1}\sum_j (B_j\xi_{0,j})^2$. 

One might expect capture to occur in interactions with many close encounters (i.e., Jacobi captures), a very close encounter, or a combination of these factors, as such events would lead to more substantial dissipation of $E_H$. Since the number of close interactions and the closeness of those encounters appear to be fractal functions of impact parameter \citep[e.g.,][]{Boekholt_2023}, we expect that the impact parameters leading to capture will occur in a fractal set (or at least a discontinuous one; see also the results of \citealp{Delaurentiis_2023} in their low-density regime). Since we randomly choose the impact parameters from a uniform distribution, the number of interactions leading to capture is probabilistic. We therefore report the ``resolution'' $\rate_{\text{res}}$, the lowest rate that we would expect to ``appear'' in our results for each value of $C_i$. We calculate these using equation (\ref{eq:rate_approx}), with the summation term set to 1 (i.e., $\rate \simeq \rate_{\text{res}}$ if only a single test with $\xi_0 \simeq r_H$ ends in capture). 

\subsection{Handling Integration Issues}
\label{sec:numerical_errs}

There is a singularity in equation  (\ref{eq:hill_eom_dimless}) as $\rho \to 0$; \texttt{scipy.ode} occasionally struggles to handle the high accelerations when the two BHs are very close. A sign of this problem is that interactions with no friction can occasionally ``lose'' enough energy through numerical effects during close encounters that they meet the conditions for capture. We experimented with two approaches to avoid this --- gravitational softening and regularization.  

We settled on softening as our benchmark technique, used for the results reported below. Another possible option would be to use \texttt{BRUTUS}, an $N$-body code that uses a combination of Burlisch--Stoer integration and exact arithmetic \citep{PZ_2014,Boekholt_2015,Boekholt_2021}.

We implemented softening of the gravitational potential by replacing $\rho^{-1}$ with $(\rho^2 + b_s^2)^{-1/2}$, where $b_s$ is some constant softening length --- in the Hill problem, it is reasonable to choose $b_s$ to be some small fraction of the mutual Hill radius $r_{\text{H}}$. We integrated batches of $10^5$ test interactions with no dynamical friction, with softening set by a range of values $b_s/r_{\text{H}} \in [10^{-9}, 10^{-1}]$. We found that all batches with $b_s/r_{\text{H}} > 10^{-8}$ yielded 0 captures, which provides a lower limit to the useful softening. In tests of interactions under GDF with dynamical friction, softened batches showed significant changes in the capture rate with $b_s$ when $b_s/r_{\text{H}} \geq 10^{-4}$. As a middle ground, we chose $b_s/r_{\text{H}} = 10^{-5}$ as our softening parameter. Although this parameter is likely large enough to alter the trajectory of certain interactions (see, e.g., Figure \ref{fig:spectra}), we expect that it will not alter the overall capture rates substantially. 

We also experimented with regularization, but we found that the batches integrated with time-regularized equations yielded similar rates of non-physical captures for $C_g = 0$ as the non-regularized, non-softened EOM. When GDF was present, the capture rates for the regularized integrations resemble those of both the softened and non-softened non-regularized integrations. This provides some reassurance that all three methods yield accurate capture rates for $C_g > 0$.

\section{Interactions and Captures}\label{sec:results}
\subsection{Individual Interactions}\label{sec:single_ints}

\begin{figure*}
    \centering
    \includegraphics[width=\textwidth]{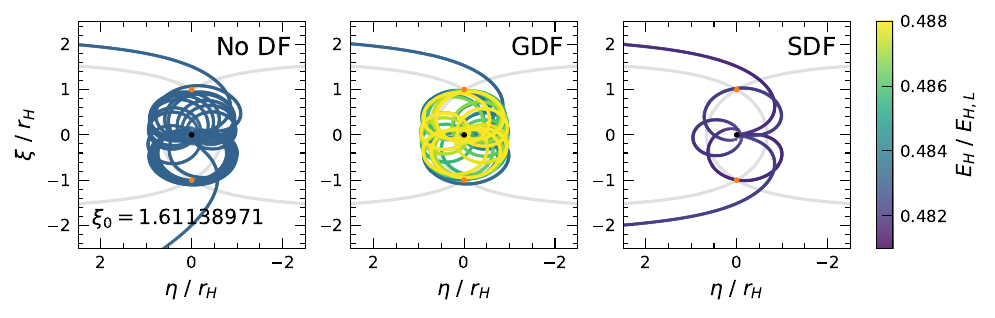}
    \caption{Three interactions with the same initial conditions ($\xi_0 = 1.61138971$, $f_e = f_I = 0$). In the plot on the left there is no dynamical friction, in the middle plot there is GDF with $C_g = 10^{-4}$, and in the right plot there is SDF with $|C_s\Gamma_2| = 10^{-4}$ (using the fiducial $\gamma=1.5$). We plot only the first $t_d=50$ time units. The black point denotes the origin (i.e., zero separation), the orange points indicate $\xi=\pm r_{\text{H}}$ (the Lagrange points $L2$ and $L1$), and the grey contours show the separatrices going through these points, at which the effective potential (i.e., the component of $E_{\text{H}}$ not dependent on velocity) is a saddle point; here $E_{\text H}=E_{\text{H},\,\text{L}}=-3^{4/3}/2$. The colored line shows the value of $\pmb{\rho}$ at timesteps of $\Delta t_d = 0.01$, with the color of the line indicating the current value of the Jacobi--Hill constant. In each panel, the system enters the frame from the  positive $\xi$- and $\eta$-direction (upper left). Though the GDF and SDF cases have equal coefficients, they evolve quite differently (see Figure \ref{fig:example_energy} for more details). The GDF case will eventually lead to capture, while the SDF case has met the conditions for escape by the end of this short simulation.}
    \label{fig:example_interaction}
\end{figure*}
\begin{figure*}
    \centering
    \includegraphics[width=\textwidth]{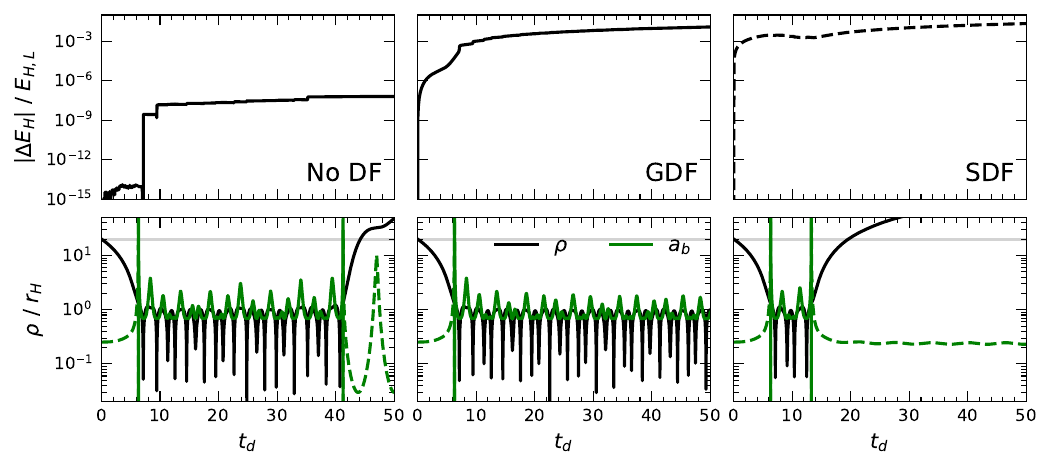}
    \caption{The Jacobi--Hill constant $E_{\text{H}}$, separation $\rho$, and binary semi-major axis $a_b$ (green lines) of the three interactions shown in Figure \ref{fig:example_interaction}, with no dynamical friction (left panels), GDF (middle panels), and SDF (right panels). The grey horizontal line in the bottom panels shows the cutoff $\rho > \rho_{\text{esc}}$ used to define escape in our simulations. The cumulative change in Jacobi--Hill constant by a given time is several orders of magnitude smaller in the absence of friction, suggesting that drift in $E_H$ from numerical errors is insignificant relative to the friction coefficients examined here. Dashed lines indicate negative values. Note that SDF yields the biggest $|\Delta E_{\text{H}}|$ in the first few time units because it acts on Keplerian orbits while GDF does not (see \S\ref{sec:preint_effects}).}
    \label{fig:example_energy}
\end{figure*}

In Figures \ref{fig:example_interaction} and \ref{fig:example_energy}, we show the properties of three example interactions between bodies approaching each other on zero-eccentricity, zero-inclination orbits; each interaction has the same impact parameter $\xi_0$, but one is integrated under no friction, one under GDF, and one under SDF. The friction coefficients are equal ($C_g = |C_s\Gamma_2| = 10^{-4}$, with $\gamma = 1.5$ for SDF).

The interaction with no friction exhibits features of a transient Jacobi capture \citep{Petit_1986,Boekholt_2023}. The bodies undergo $N_e = 20$ close encounters (counted as minima of $\rho$ while $\rho < r_H$). We cut off each of these integrations at $t_d = 50$ to keep the figures legible, but we know from longer integrations that the interaction under GDF would eventually lead to permanent capture, according to the conditions set in the previous section. The interaction under SDF does not lead to permanent capture, and the bodies drift apart after only one close encounter.

From the left panel of Figure \ref{fig:example_energy}, we see that there are still small, non-physical changes to $E_{\text{H}}$ during close encounters, even in the absence of friction. In this integration we have used a softening parameter $b/r_{\text{H}} = 10^{-5}$, which we believe has eliminated the worst of these numerical errors (see discussion in \S\ref{sec:numerical_errs}). These numerical ``kicks'' in $E_{\text{H}}$ are many orders of magnitude smaller than the change from GDF/SDF during similarly close encounters, and in the absence of friction the Jacobi--Hill constant is generally conserved to better than one part in $10^6$. With this level of energy conservation, we expect that the capture rate evaluated from large batches of integrations will be accurate, even though individual integrations with multiple close encounters might not be.

Figure \ref{fig:example_energy} highlights one of the key differences between GDF and SDF --- they act differently on the azimuthal or $\hat{\pmb{\eta}}$ component of the relative velocity vector $\pmb{\rho}'$. As this is, initially, the only non-zero component of the vector for incoming particles on circular orbits, this difference strongly affects how much the Jacobi--Hill constant $E_{\text{H}}$ will change prior to the first close encounter. Under SDF, the cumulative change to the Jacobi--Hill constant $\Delta E_{\text{H}} / E_{\text{H},\,0} \equiv (E_{\text{H},\,t_d} - E_{\text{H},\,0})/E_{\text{H},\,0}$ is already $\gtrsim 10^{-3}$ as the bodies near their first close approach; while under GDF, the change in $E_{\text{H}}$ in the leadup to the first close approach is over two orders of magnitude smaller.

\subsection{Initially Circular Orbits}\label{sec:spectra}

\begin{figure}
    \centering
    \includegraphics[width=.47\textwidth]{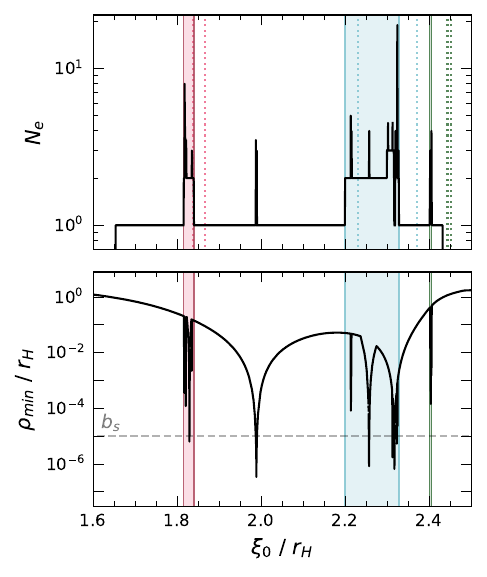}
    \caption{Spectra of various interaction properties (in the absence of dynamical friction) across a portion of the tested range of impact parameter $\xi_0$. (Top) Number of close encounters $N_e$ during an interaction. Dotted lines border the regions of Jacobi capture determined by \citet{Boekholt_2023} in their ``BH+'' simulations --- shaded regions denote these regions based on our data. (Bottom) Separation at closest encounter, with the softening parameter used in our integrations noted as a dashed grey line. Note that ranges of impact parameter in which $\rho_{\text{min}} < b_{\text{s}}$ correspond to small regions in the top panel that exhibit sharp peaks in $N_e$. These peaks are presumably artifacts of the softening  (cf. Figure 4 of \citealp{Boekholt_2023}). Our overall capture rates $\rate(C_i > 0)$ do not appear to be significantly impacted by these artifacts (see discussion in \S \ref{sec:numerical_errs}).}
    \label{fig:spectra}
\end{figure}

\begin{figure*}
    \centering
    \includegraphics[width=\textwidth]{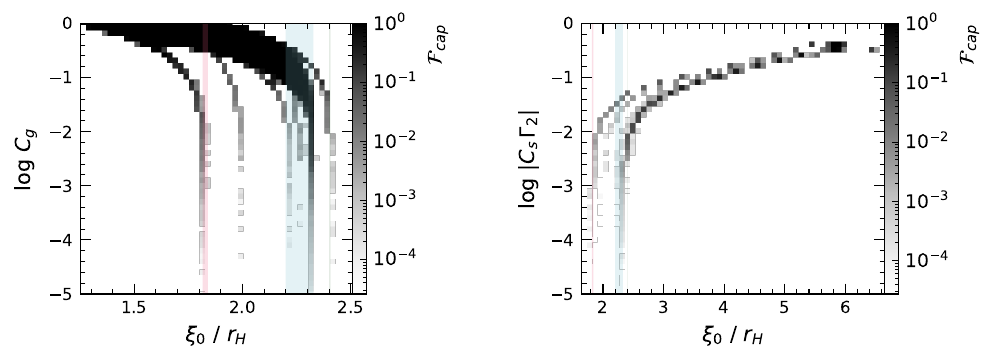}
    \caption{Fraction of interactions leading to capture, $\mathcal{F}_{\text{cap}}$, in bins of impact parameter ($\xi_0$) vs.\ friction coefficient ($C_i$) for both GDF (left) and SDF (right). The approximate Jacobi capture regions from the dissipation-free spectrum of the number of close encounters $N_e$ (Figure \ref{fig:spectra}) are shown as shaded regions of $\xi_0$.}
    \label{fig:2d_hists}
\end{figure*}
\begin{figure*}
    \centering
    \includegraphics[width=\textwidth]{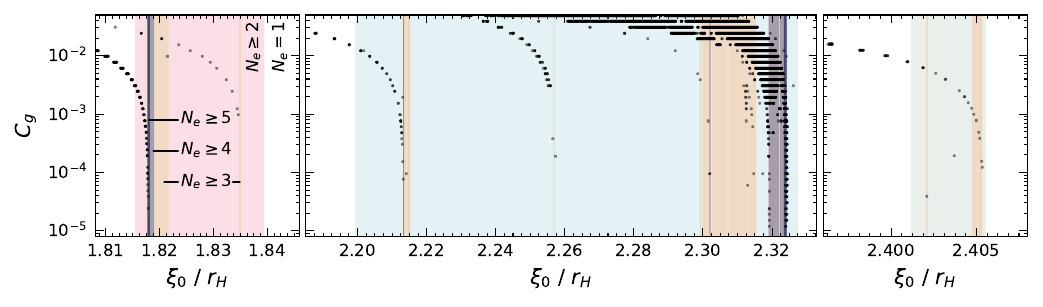}
    \caption{Impact parameters leading to BBH formation across a range of $C_g$ (black dots), with each panel zoomed in to show a Jacobi capture region (widest shaded region in each) from Figures \ref{fig:spectra} and \ref{fig:2d_hists}. Impact parameter ranges with $\geq 3$ close encounters ($N_e$) in the friction-free tests are shaded in beige, regions with $N_e \geq 4$ are shaded in light purple, and regions with $N_e \geq 5$ are shaded in dark purple. For $C_g \ll 10^{-3}$, the impact parameters leading to BBH formation are tightly related to those yielding long-lived Jacobi capture in the friction-free tests.}
    \label{fig:gdf_zoomed}
\end{figure*}

Here, we describe some properties of the interactions as a function of the initial impact parameter $\xi_0$ (which we refer to as ``spectra'') and the friction coefficient $C_i$. We focus on class (1) interactions (i.e., those beginning with circular, coplanar orbits). 

In Figure \ref{fig:spectra}, we show spectra of the number of encounters during an interaction $N_e$ and the separation at closest approach $\rho_{\text{min}}$ when there is \textit{no} dynamical friction, which can be compared to similar plots in the slightly dissimilar setup of \citet{Boekholt_2023} (see below for more).

We track $N_e$ for a given encounter by incrementing a counter every time $\rho$ starts to increase while $\rho < r_{\text{H}}$. To find the minimum separation, we record $\rho$ at every integrator-specified step (rather than the user-defined $\Delta t_d = 0.1$, which occasionally sampled too coarsely) and keep the smallest separation on record during the course of an interaction. The spectra show the median parameter value in bins of width $\delta\xi_0/r_H = 10^{-4}$. 

The spectra of $N_e$ and $\rho_{\text{min}}$ can be compared to Figures 4 and 5 of \citet{Boekholt_2023}. There, the authors examine the Jacobi (transient) capture of two particles of \textit{different} masses on nearby circular orbits. They present high-resolution spectra of $N_e$ and $\rho_{\text{min}}$, identifying three ranges of impact parameter in which Jacobi capture is common --- we refer to these as ``Jacobi capture regions'' --- which are related to the ``transitional'' zones identified in \citet{Petit_1986}. (Boekholt et al.\ also analyze the fractal nature of capture within these regions, a level of detail that we do not attempt to capture.) The bounds of the three Jacobi capture regions identified from their integrations are shown as dotted lines in the top panel in Figure \ref{fig:spectra}. We make our own estimate of the Jacobi capture regions for \textit{equal-mass} bodies by identifying the main departures from the baseline $N_e = 1$. These are somewhat different from those of \citet{Boekholt_2023}, probably because Boekholt et al.\ used particles with a mass ratio of $0.78$ rather than unity. Our use of a softened potential may also contribute to the differences. 

In Figure \ref{fig:2d_hists}, we show the regions of $\xi_0$--$C_i$ space that yield captures in class (1) interactions under both GDF and SDF. Here, $\mathcal{F}_{\text{cap}}$ denotes the fraction of interactions in a given bin of this parameter space that led to capture. For weak dynamical friction in gaseous systems, $C_g \lesssim 10^{-2}$, the regions of $\xi_0$ that yield captures are relatively independent of $C_g$ and consistent with the regions of Jacobi capture identified in the frictionless case (Figure \ref{fig:spectra}); at higher $C_g$, these capture-inducing regions shift inwards and expand. The regions of capture under high friction coefficients are similar to those found in \citet{Qian_2024} and \citet{Delaurentiis_2023}, which we discuss more thoroughly in $\S$\ref{sec:compare}. Though we do not distinguish prograde and retrograde captures in this figure, note that prograde captures occur in GDF simulations with $C_g > 3\times10^{-2}$ in the same regions found by \citet{Qian_2024} --- we also find a very small percentage of captures with $\xi_0 \simeq 2.3r_H$ and $C_g \simeq 3\times 10^{-3}$ to be prograde, but otherwise find no prograde captures below $C_g = 3\times10^{-2}$. We discuss prograde capture rates further in \S\ref{sec:cap_rates}.

The distribution of captures under SDF is clearly impacted by pre-interaction friction (as discussed in \S\ref{sec:preint_effects}). For $|C_s\Gamma_2| \lesssim 10^{-3}$, the impact parameter ranges leading to capture are consistent with the frictionless Jacobi capture regions (as was the case with weak GDF); however, at higher values of $|C_s\Gamma_2|$, the capture region shifts rapidly to larger impact parameters. This is because SDF causes the two bodies to drift radially at different rates --- the outer body drifts more quickly, so large impact parameters at large azimuthal separations (recall that we start these integrations with $\eta_0 = 20\,r_H$) have shrunk by the time the bodies reach each other. As discussed in the fourth paragraph of \S\ref{sec:preint_effects}, we do not trust our circular-orbit ICs for SDF in this friction regime. 

Lastly, in Figure \ref{fig:gdf_zoomed}, we zoom in to the Jacobi capture regions to examine the interplay between long-lived interactions in friction-free tests and permanent BBH formation under GDF. Here, we identify different lengths of Jacobi capture (i.e., different ``heights'' in the spectrum of close encounters $N_e$ in Figure \ref{fig:spectra}) with different colored shadings. It is clear that under weak friction ($C_g \ll 10^{-3}$), instances of BBH formation are tightly related to impact parameters with long-lasting Jacobi capture in the friction-free test; however, for $C_g \gtrsim 10^{-3}$, this correlation is broken (likely by dissipation prior to the first close encounter). As we did not integrate the same number of interactions at each $C_g$, not much should be made of the raw number of BBHs formed at any given $C_g$ --- Figure \ref{fig:gdf_zoomed} is meant to display a general trend rather than to draw focus to individual instances of BBH formation.

\subsection{Eccentric/Inclined Orbits}\label{sec:prop_noncirc}

\begin{figure*}
    \centering
    \plottwo{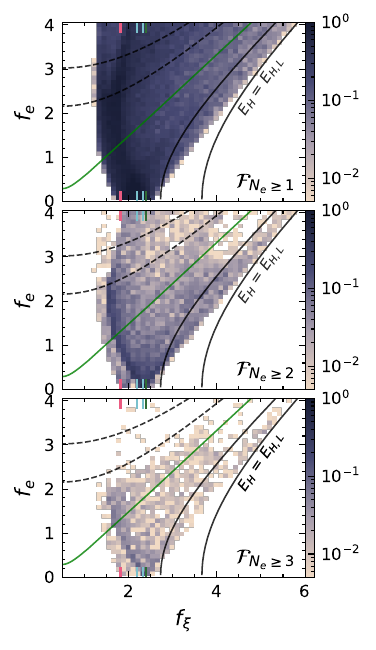}{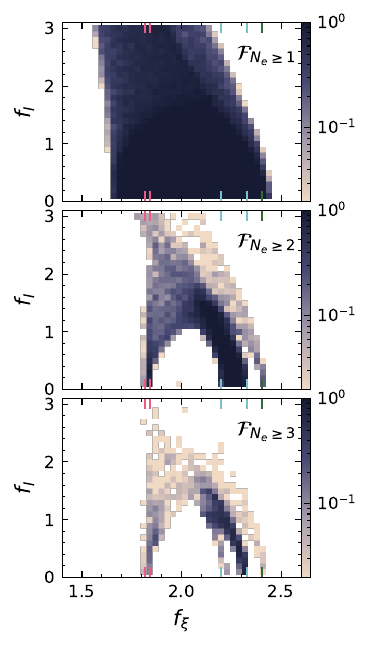}
    \caption{Distribution of encounters across (left) impact parameter--eccentricity space (with inclination $f_I = 0$) and (right) impact parameter--inclination space (with eccentricity $f_e = 0$). There is no friction in these simulations. The color of each cell shows the fraction of interactions with initial conditions in that cell that experienced at least one close encounter (top), at least two close encounters (middle), and at least three close encounters (bottom). Recall that we define ``close encounters'' as pericenter passages with $\rho<r_H$. These plots show that the narrow ranges of impact parameters that yield $N_e \geq 2$ (i.e., Jacobi capture) for circular, flat initial orbits broaden at non-zero  eccentricity and inclination. Colored ticks at the top and bottom of each subplot show the boundaries of the Jacobi capture regions for initially circular and flat orbits. We show linearly spaced contours of the initial Jacobi--Hill constant $E_{\text{H}}$ across this space of initial conditions between $E_{\text{H}} = E_{\text{H,L}}$ (bottom, solid line) and $E_{\text{H}} =-E_{\text{H,L}}$ (top, dashed line). Recall that $E_{\text{H,L}} \equiv -3^{4/3}/2$ the value of the Jacobi--Hill constant for a stationary system at the first and second Lagrange points. The $E_{\text{H}}=0$ contour is highlighted in green. Note that all orbits to the left of the $E_{\text{H}} = E_{\text{H,L}}$ contour have $E_H > E_{\text{H,L}}$ (and thus have enough ``energy'' reach the origin).}
    \label{fig:ne_fefi}
\end{figure*}

\begin{figure*}
    \centering
    \includegraphics[width=\textwidth]{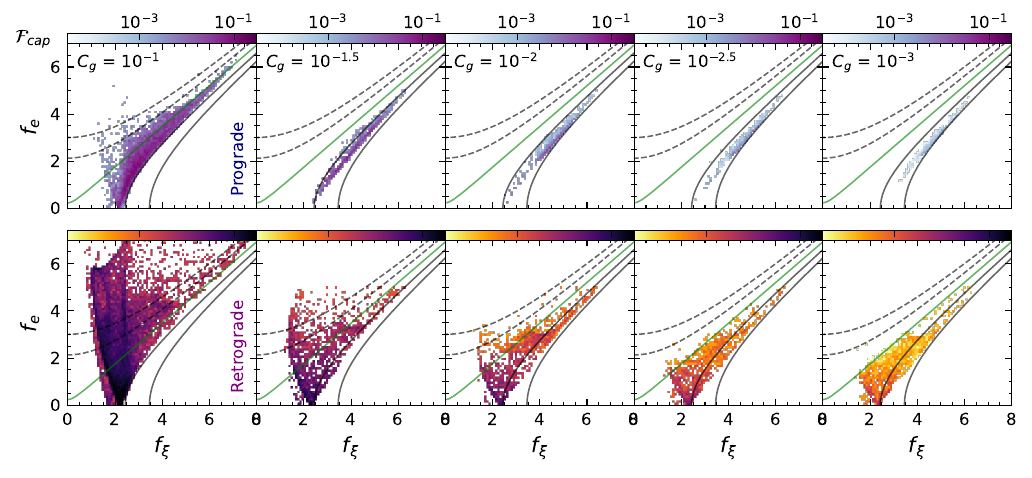}
    \caption{Capture probabilities $\mathcal{F}_{\text{cap}}$ for interactions between bodies on initially eccentric orbits across impact parameter--eccentricity space (with $f_I=0$). The top row shows prograde captures (blue-green shading) for a range of friction parameters $C_g$; the bottom row shows retrograde captures under the same conditions (purple-orange shading). All colorbars share the same scaling. We show the same contours of initial $E_{\text{H}}$ as in Figure \ref{fig:ne_fefi} --- i.e., linearly spaced between $E_{\text{H}} = E_{\text{H,L}}$ (bottom, solid black line) and $E_{\text{H}} =-E_{\text{H,L}}$ (top, dashed black line), with $E_{\text{H,L}}$ the value of the Jacobi--Hill constant for a stationary system at the first and second Lagrange points. The $E_{\text{H}}=0$ contour is highlighted in green. Recall that $E_{\text{H,L}}$ is negative; points above the green contour have $E_{\text{H}} > 0$, while points below the contour have $E_{\text{H}} < 0$.}
    \label{fig:fcap_fe_2dhist}
\end{figure*}
\begin{figure*}
    \centering
    \includegraphics[width=\textwidth]{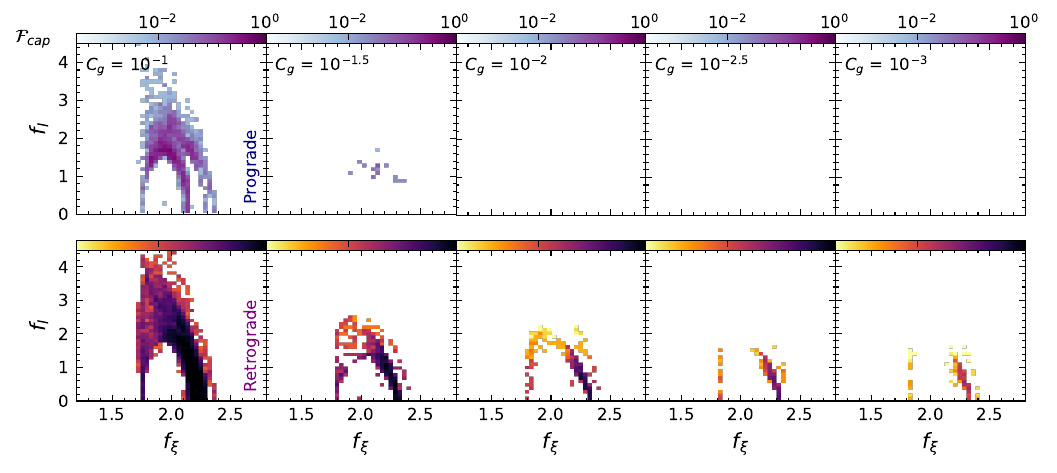}
    \caption{Equivalent to Figure \ref{fig:fcap_fe_2dhist}, but across impact parameter--inclination space (with $f_e=0$). We found no prograde captures with $C_g \leq 10^{-2}$.}
    \label{fig:fcap_fi_2dhist}
\end{figure*}

We now turn our attention to class (2) interactions. Each of these has five characteristic parameters --- the initial difference in semimajor axes $f_\xi$, the amplitudes of the radial/azimuthal and vertical epicycles in units of the Hill radius ($f_e$ and $f_I$, respectively), and the epicyclic phases $\delta_e$ and $\delta_I$\footnote{We also note that (\textit{i}) $\eta_0$ is also relevant, but in the frictionless case its effect is degenerate with the dependence on epicyclic phase, and (\textit{ii}) all of our class (2) interactions have either non-zero eccentricity and zero inclination or vice versa.} --- making the range of possible encounters far more complex than for class (1) cases.

For a given value of $f_e$ and $f_I$ we can find a spectrum of the number of close encounters $N_e$ (defined as the number of times a pair of bodies reaches periapsis with $\rho < r_H$ during an interation). These will be analogous to the spectra we found in the class (1) case $f_e=f_I=0$, except that now the spectra are averaged over the random phases $\delta_e$ and $\delta_I$. To do this,  we integrate $10^4$ interactions without friction, at linearly spaced increments $\Delta f_e = \Delta f_I = 0.1$, over the ranges $f_e \in [0.1,4.0]$ and $f_I \in [0.1,3.0]$. Each interaction has a randomly drawn $f_\xi \in [0.5,6]$ for eccentric interactions and $f_\xi \in [0.5,3.5]$ for inclined ones, along with a random phase $\delta_e$ or $\delta_I$ $\in[0,2\pi]$. 

In Figure \ref{fig:ne_fefi}, we show the Jacobi capture regions across $f_\xi$--$f_e$ and $f_\xi$--$f_I$ space. In the top subplot of each figure, we show the distribution of interactions with at least one close encounter (compare Figure \ref{fig:spectra} for initially circular orbits); recall that $\xi_0/r_H = f_\xi$ when $f_e = 0$. In the bottom two subplots, we look at the distribution of interactions with at least two, then at least three close encounters. 

We first focus on eccentric interactions. As $f_e$ grows from 0, the range of impact parameters leading to Jacobi capture spreads out in either direction from the relatively narrow bands present in class (1) tests. The regions with $N_e \geq 2$ are similar to those with $N_e \geq 1$. Above $f_e \sim 2$, the regions with $N_e \geq 1$ or $N_e\geq 2$ are weighted towards small impact parameters, while those with $N_e\geq 3$ are weighted toward large impact parameters. 

For inclined orbits with $f_I \lesssim 1$, the Jacobi capture regions are similar to those in the circular, flat case --- there are three distinct regions with $N_e \geq 2$. As the inclination gorws, the second and third of these regions shift gradually to smaller impact parameters. For $f_I \gtrsim 1$, the capture regions merge into a single region, which continues to shift toward smaller impact parameters as $f_I$ increases. 

It is important to understand more fully the dependence of the Jacobi capture regions on eccentricity and inclination. We shall not attempt to do so in detail in this paper. However, as a first step we show in Figures \ref{fig:ne_fefi} and \ref{fig:fcap_fe_2dhist}  contours of the Jacobi--Hill constant $E_H$ (eq.\ \ref{eq:E_H}) as a function of $f_\xi$ and $f_e$. All orbits for which Jacobi capture is possible have initial $E_{\text{H}} > E_{\text{H,L}}$, the value of the Jacobi--Hill constant for a stationary system at the first and second Lagrange points, as is required for the system to have enough ``energy'' to overcome the effective potential barrier and have a close encounter with the origin. As $f_e$ grows, the maximum $f_\xi$ yielding Jacobi capture seems to trace the $E_{\text{H}} = E_{\text{H,L}}$ contour --- encounters with energy large enough for a close encounter often have one. In contrast, 
the minimum $f_\xi$ boundary of the capture region at a given $f_e$ does not align with any contour of $E_{\text{H}}$, and the boundary seems to be less clearly defined. 

In Figures \ref{fig:fcap_fe_2dhist} and \ref{fig:fcap_fi_2dhist}, we show the distribution of permanent binary formation (based on the capture conditions decribed earlier) across $f_\xi$--$f_e$ and $f_\xi$--$f_I$ space, for several values of the friction coefficient $C_g$. In general, the regions of permanent capture are roughly consistent with the regions of Jacobi capture, especially at low $C_g$ ($\lesssim 10^{-2}$). This behavior is consistent with our class (1) findings. We conclude that the formation of long-lived BBHs under the influence of weak dynamical friction requires that the single BHs experience Jacobi capture with at least a few close encounters; with stronger dynamical friction, interactions that would otherwise undergo only one or two close encounters may also lead to the formation of a binary.

Although we find only a few prograde captures from flat, circular (class 1) encounters with $C_g < 3\times10^{-2}$, such captures are more common if the initial orbits are eccentric. For $C_g < 10^{-1}$, these captures trace out a narrow band of $f_\xi$-$f_e$ space. 

For initially inclined orbits, prograde captures form $\sim 20\%$ of binaries when $C_g = 10^{-1}$ and $\sim 1\%$ when $C_g = 10^{-1.5}$; we do not find any prograde captures for smaller values of $C_g$. We discuss the orientation of the formed BBHs more fully in \S\ref{sec:properties}.

\section{Formation Rates}\label{sec:cap_rates}
 
\subsection{Initially Circular Orbits}\label{sec:circular}

\begin{figure}
    \centering
    \includegraphics[width=.47\textwidth]{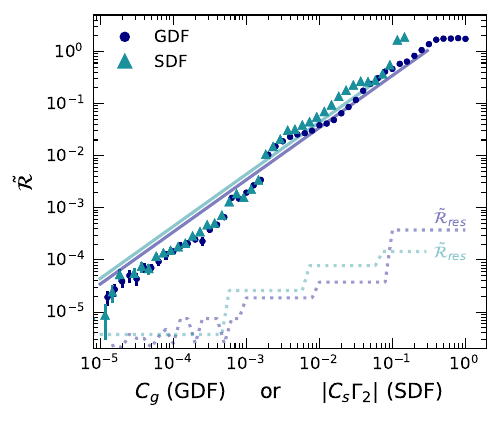}
    \caption{The scaled capture rate $\rate$ under the influence of GDF (navy) and SDF (teal), as a function of the friction coefficients $C_g$ and $C_s$. Each point at a particular coefficient (markers with error bars) has a corresponding ``resolution'' (dotted line), which marks the minimum detectable rate given how many interactions have been simulated with that coefficient (see \S\ref{sec:calculate}). Lastly, we plot lines using the intercept parameters of our G/SDF ``full'' fits with the power law parameter $\beta$ fixed to $1$ (solid lines; see Table \ref{tab:fits} and text for more on these fits).}
    \label{fig:rates_class1}
\end{figure}

In Figure \ref{fig:rates_class1}, we show the scaled binary formation rate per unit area $\rate(C_i)$ for initially circular and flat orbits ($f_e = f_I = 0$) under GDF and SDF, across a range of the friction coefficients $C_g$ and $|C_s\Gamma_2|$. For SDF, we present results as a function of $|C_s\Gamma_2|$ rather than $C_s$ alone because the former more accurately describes the friction coefficient acting on a system when the bodies are widely separated (see \S\ref{sec:preint_effects} for more). Error bars come from equation (\ref{eq:rate_error}) and represent the standard deviation of the Monte Carlo integration. 

Figure \ref{fig:rates_class1} suggests that we have reliable rate estimates for most coefficients in the range $10^{-5} \lesssim C_g \lesssim 1$ and $10^{-5} \lesssim |C_s\Gamma_2| \lesssim 10^{-1}$. In general, we aimed to attain  more than 5 captures for every coefficient, although the data points with the smallest friction coefficients shown under GDF and SDF represent only 4 and 1 capture(s), respectively.

For both GDF and SDF, the capture rates for coefficients $C_g,|C_s\Gamma_2| \lesssim 10^{-3}$ appear to be roughly linear functions of $C_g$ or $|C_s\Gamma_2|$. The rates above $C_g,|C_s\Gamma_2| \sim 2\times 10^{-3}$ appear to be inflated relative to the rates at lower friction coefficients, by about a factor of 
two. The rate $\rate_{\text{GDF}}$ continues to increase (roughly) linearly with $C_g$ until $C_g \sim 0.4$, at which point its increase flattens to zero. We caution again that the physical approximations used in our model become less realistic and less self-consistent at such high values of the friction coefficient, as discussed in \S\ref{sec:preint_effects}.

The rate $\rate_{\text{SDF}}$ increases up to $|C_s\Gamma_2| \sim 10^{-1}$, although we refer the reader back to the warnings in \S\ref{sec:preint_effects} and Figure \ref{fig:2d_hists} --- at these large friction coefficients our physical assumptions and ICs are not very realistic. 

We have experimented with fitting the capture rates for circular, flat initial orbits (class 1) to the simple relationships
\begin{align}\nonumber
    \rate_{\text{GDF}}(C_g) &=  \alpha_g C_g^{\beta_g}
    \\ \label{eq:rate_model}
    \text{and}\quad
    \rate_{\text{SDF}}(C_s;\Gamma_2) &=  \alpha_s |C_s\Gamma_2|^{\beta_s},
\end{align} 
where $\alpha_i$ and $\beta_i$ are free parameters. We estimate these parameters by a linear regression on our log-scaled $\rate$ and $\rate_{\text{err}}$ data, using \href{https://docs.scipy.org/doc/scipy/reference/generated/scipy.optimize.curve_fit.html}{\texttt{scipy.optimize.curve\_fit}}. For these regressions, we only consider coefficient values with more than 5 captures. We report our best-fit parameters in Table \ref{tab:fits}.
\begin{table}[]\label{tab:fits}\centering
    \begin{tabular}{| c | c c | c |}
    \multicolumn{1}{c}{$\rate = \alpha X^\beta$, with} & \multicolumn{2}{c}{$\beta$ free} & \multicolumn{1}{c}{$\beta = 1$}\\
    \hline
    Data & $\beta_i$ & $\alpha_i$ & $\alpha_i$ \\
    \hline
    \hline
    GDF, full & $1.13\pm0.02$  & $6.15\pm0.60$ & $3.45\pm0.26$\\
    SDF, full & $1.21\pm0.03$  & $13.23\pm2.31$ & $4.41\pm0.46$\\
    \hline
    GDF, high & $0.98\pm0.02$ & $4.06\pm0.37$ & $4.32\pm0.16$ \\
    SDF, high & $0.94\pm0.03$ & $4.96\pm0.79$ & $6.28\pm0.26$\\
    \hline
    GDF, low & $1.06\pm0.04$ & $2.61\pm1.15$ & $1.60\pm0.10$\\
    SDF, low & $1.02\pm0.04$ & $2.08\pm0.77$ & $1.81\pm0.09$\\
    \hline
    \end{tabular}
    \caption{Best-fit parameters and standard deviations for various fits of the data in Figure \ref{fig:rates_class1} to the model (\ref{eq:rate_model}). ``Full'' considers data in the domains $10^{-5}\leq C_g\leq0.3$ and $10^{-5}\leq|C_s\Gamma_2|\leq7.5\times10^{-2}$; ``high'' and ``low'' consider the high- and low-friction regimes of these domains, split at $C_g,|C_s\Gamma_2| = 1.8\times10^{-3}$. The second and third columns give best fits with both $\alpha_i$ and $\beta_i$ as free parameters, while the last column gives best fits of $\alpha_i$ with $\beta_i$ set to unity.}
\end{table}    

We first perform a fit considering $C_g \in [10^{-5},0.3]$ and $|C_s\Gamma_2| \in [10^{-5},7.5\times10^{-2}]$ --- this is the complete set of reported values, neglecting coefficients for which pre-encounter dissipation may be important (as described in \S\ref{sec:preint_effects}), and constitutes our ``full'' fitting data. Close inspection of Figure \ref{fig:rates_class1} shows an apparent break at $C_g, |C_s\Gamma_2| \simeq 1.8\times10^{-3}$. While we do not have a physical explanation for this break --- we recognize that it may be an undiscovered numerical error --- we fit lines separately to the data on either side of the break to avoid a possible bias in the slopes $\beta_i$.  We do this over the same overall ranges of data from the ``full'' fit, considering coefficients $< 1.8\times10^{-3}$ (``low'') and $> 1.8\times10^{-3}$ (``high'') separately. Finally, we perform the same six fits (full, high, and low regimes for both GDF and SDF) with $\beta_i$ set to $1$ --- that is, \textit{assuming} that the formation rate is linear in the friction coefficient.

As reported in Table \ref{tab:fits}, neither of the fits to the full data, with $\beta_i$ a free parameter, are consistent with $\beta=1$ --- they both yield a slightly steeper slope, presumably reflecting the discontinuity at coefficient $\simeq 1.8\times10^{-3}$. All fits to the data on one side or the other of the discontinuity are consistent with $\beta_i=1$, to within $2\sigma$ error. 

We conclude that the relationship between capture rate and friction coefficient is approximately linear, $\rate \propto C_g,|C_s\Gamma_2|$, over at least four orders of magnitude in friction coefficient. The main systematic deviations from linearity are likely to arise from (\textit{i}) the discontinuity of unknown origin and (\textit{ii}) our crude approach to modeling both GDF and SDF. We assume that this linear relationship continues to even lower values of $C_g$ and $C_s$. Our results differ from those of \citet{Qian_2024}, who found no captures below $C_g \simeq 3\times10^{-2}$. We expect that this finding is a result of their sparse sampling in impact parameter. See \S\ref{sec:compare} for further comparison of our results to this and other recent works.

We may summarize this discussion in the relationships
\begin{align}
    \rate_{\text{GDF}} / C_g &\simeq 3.5 \pm 0.3 \nonumber\\
    \text{and}\quad
    \rate_{\text{SDF}} / |C_s\Gamma_2| &\simeq 4.4 \pm 0.5.
    \label{eq:bestfit}
\end{align}
These are shown as solid lines in Figure \ref{fig:rates_class1}.

These simple rules can be used to estimate binary formation rates in any scenario for which (\textit{i}) the approximations in our derivations of GDF and SDF are valid and (\textit{ii}) $C_g \lesssim 10^{-1}$ and $C_s \lesssim 10^{-2}$. The deviations from these fits, and the errors due to the approximations we have made in deriving them, are much smaller than the uncertainties arising from our limited understanding of the properties of AGN disks. Rates set by these linear fits would be suitable for use in, e.g., population synthesis models of BBHs like that of \citet{Tagawa_2020}.

\begin{figure*}
    \centering
    \includegraphics[width=\textwidth]{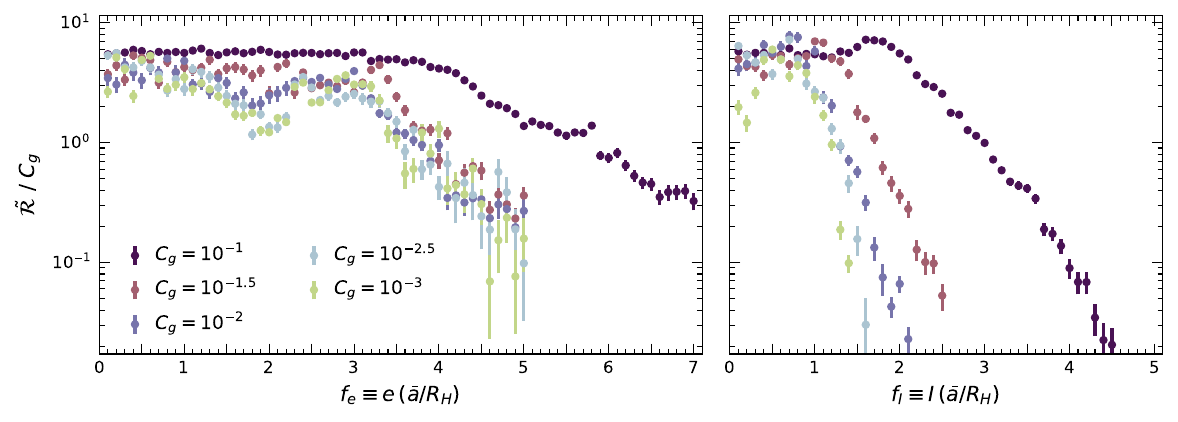}
    \caption{Capture rates $\rate$ for interactions under GDF when the two bodies have a mutual epicyclic amplitude $f_e$ (radial/azimuthal, corresponding to eccentricity) or $f_I$ (vertical, corresponding to inclination) at the initial separation $\eta = 20r_H$. We show capture rates for different values of the GDF strength $C_g$ across a range of these amplitudes, which can be related to the initial eccentricity and inclination of the single BHs as described in \S\ref{sec:ICs} and \S\ref{sec:calculate}. Recall that $\bar{a}$ is the distance from the two bodies of interest to the central body, and $R_H$ is their mutual Hill radius in physical units (eq.\ \ref{eq:hilldef}).}
    \label{fig:rate_ei}
\end{figure*}

\subsection{Eccentric/Inclined Orbits}

In Figure \ref{fig:rate_ei}, we present estimates of the scaled capture rate for interactions between bodies on initially eccentric or inclined orbits. Generally, for $f_{e}, f_{I} \lesssim 1$ the rate $\rate$ is roughly independent of $f_e$ or $f_I$ and consistent with the value we reported earlier for circular/flat (class 1) integrations. For $f_e,f_I\gtrsim 1$ the rate decays rapidly with increasing epicyclic amplitude.

With $C_g \leq 10^{-1.5}$, the scaled, normalized formation rate $\rate/C_g$ as a function of $f_e$ appears to be almost independent of $C_g$ --- the data for these friction strengths in the left panel of Figure \ref{fig:rate_ei} roughly coincide --- and the rate begins to decay at $f_e \approx 3$. With $C_g = 10^{-1}$, the value of $\rate/C_g$ shows less variation as a function of $f_e$ than under weaker friction, and the decay is slower and begins at $f_{e} \approx 4$. For all friction strengths tested, it is unclear whether the decay in $\rate/C_g$ at high $f_e$ is exponential or logarithmic (curves fit under both models return similar coefficients of determination $R^2$).

The profiles of $\rate/C_g$ versus the initial vertical epicyclic amplitude $f_I$ follow similar trends. The rate begins to decay at $f_I \approx 1$ for $C_g \leq 10^{-2}$; for $C_g = 10^{-1.5}$, $\rate/C_g$ begins to decay at $f_I \approx 1.5$, and for $C_g = 10^{-1}$, the decay does not begin until $f_I \approx 2$. The profile for $C_g = 10^{-1}$ exhibits a small but significant bump at $f_{I} \simeq 1.7$,  which corresponds to the peak prograde capture rate (see Figure \ref{fig:fcap_fi_2dhist}) --- no other profiles in inclination exhibit such bumps because prograde captures are almost entirely absent at smaller values of $C_g$. 

We note that our conclusion of the previous subsection --- that the capture rate $\rate$ is a linear function of the friction coefficient $C_g$ --- remains roughly true when the eccentricity and inclination are non-zero, at least for $C_g \lesssim 10^{-1.5}\simeq 0.03$.

\subsection{Properties of Formed Binaries}\label{sec:properties}

\begin{figure}[]
    \centering
    \includegraphics[width=.47\textwidth]{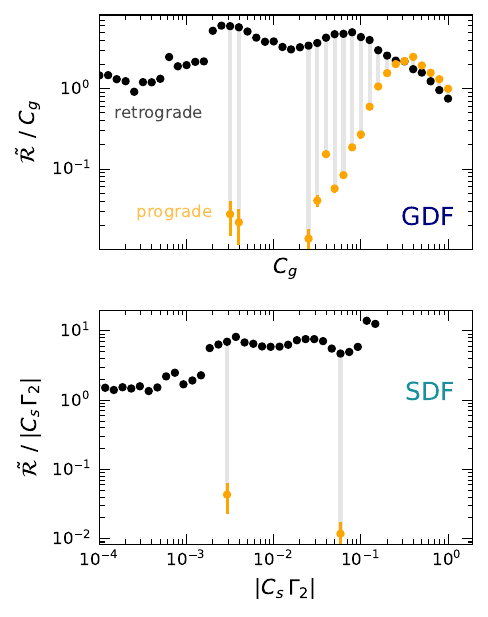}
    \caption{Scaled capture rate of prograde (orange) and retrograde (black) BBHs under GDF (top) and SDF (bottom). Grey lines connect data for the same friction coefficient for which the prograde capture rate is non-zero. The vertical axis is scaled by friction strength for readability. Under both forms of friction, most recorded captures are retrograde, and all are retrograde for sufficiently small friction coefficients.}
    \label{fig:prograde}
\end{figure}

\begin{figure*}
    \centering
    \plottwo{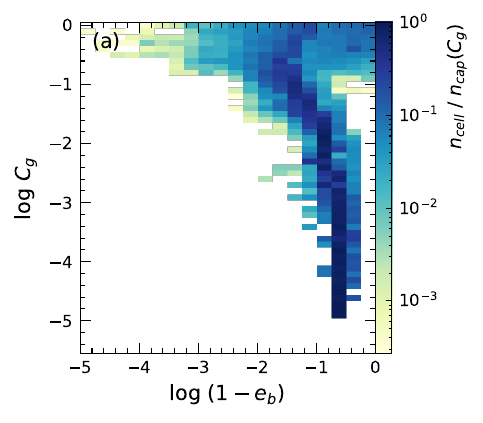}{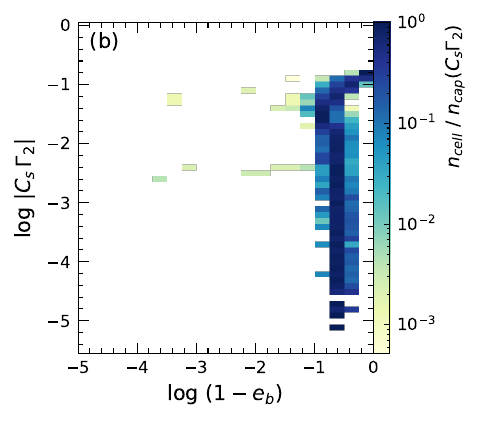}
    \plottwo{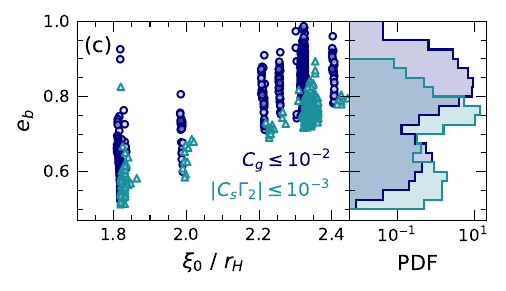}{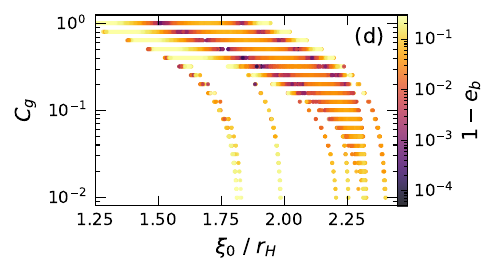}
    \caption{Eccentricities $e_b$ of the BBHs at the time of formation (as defined by the criteria in \S\ref{sec:capture}). {(a)} Histogram of $e_b$ vs. $C_g$, where the number of BBHs in each cell is scaled by the number of total captured BBHs at that coefficient. The mode of the eccentricity distribution increases with $C_g$. Note that the horizontal axis shows $\log(1-e_b)$ since most of the eccentricities are near unity. {(b)} The same for SDF. {(c)} Binary eccentricity $e_b$ vs.\ impact parameter $\xi_0$ for binaries formed under SDF and GDF with $C_i \leq 10^{-2}$. Blue circles denote GDF captures while teal triangles denote SDF. Each distinct Jacobi-capture region (distinct vertical distributions) yields  a different distribution of $e_b$. The right subplot shows histograms of $e_b$ for the same captures, showing a clear bimodality under both types of dynamical friction. {(d)} Binaries formed under GDF across a range of impact parameter $\xi_0$ and friction coefficient $C_g \geq 10^{-2}$; color denotes the value of $\log(1-e_b)$ at the moment of capture. }
    \label{fig:eb_plots}
\end{figure*}

\begin{figure*}
    \centering
    \includegraphics[width=\textwidth]{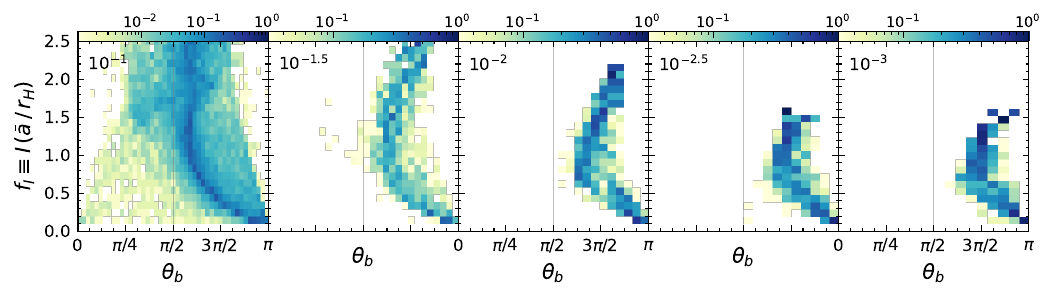}
    \caption{Inclinations $\theta_b$ of BBHs at the time of formation, for various initial inclinations $f_I$ and friction coefficients $C_g$ (values in the top left corner of each panel). Every plot spans $\theta_b \in [0,\pi]$, but we do not show the horizontal axis tick labels for all plots to reduce clutter.}
    \label{fig:inc_hists}
\end{figure*}

The properties of these dynamically formed binaries are of interest; here, we report on their inclinations and eccentricities. The values reported here are those at the moment when the capture criteria are first met (i.e., $E_H < E_{H,L}$ and $a_b < 0.05 r_H$) and these properties would continue to evolve through a variety of effects, including features of G/SDF not adequately captured in our simple models. See item 6 of the arguments in \S\ref{sec:state} for references on post-formation evolution.

First, we ask whether newly formed BBHs are prograde or retrograde. This is determined by the orientation of the specific angular-momentum vector $\mathbf{l}_b$. If the $\hat{\pmb{\zeta}}$-component of this vector is positive (i.e., if $\xi\eta' > \eta\xi'$), the BBH is prograde --- if it is negative, the BBH is retrograde.

In Figure \ref{fig:prograde}, we show the scaled formation rate of prograde binaries under both GDF and SDF for class (1) orbits. For GDF, in agreement with \citet{Qian_2024}, we find that prograde BBH formation is inefficient at low friction coefficients, although in contrast to that paper we do not find strong evidence that the prograde BBH formation rate drops to zero. In particular, we see a small but non-zero prograde capture rate at friction coefficients as small as $C_g=3\times 10^{-3}$. For even smaller $C_g$ it becomes prohibitively expensive to obtain a statistically significant sample of prograde captures. For SDF, we find that prograde formation is inefficient at all values of $|C_s\Gamma_2|$, although we find prograde BBHs at $|C_s\Gamma_2|$ as small as $ = 3\times10^{-3}$. These could reflect either a real, but small, prograde capture rate or occasional integration errors. Either way, prograde BBH formation under SDF appears to be very rare relative to retrograde formation. We discuss the implication of this result for the expected effective spins of BBH mergers in AGN disks briefly at the end of our \S\ref{sec:compare}.

We also report the distribution of eccentricity $e_b$ for newly formed BBHs. In Figure \ref{fig:eb_plots}, we show how the typical value of $e_b$ varies with friction coefficient and $\xi_0$. In the low-friction regime ($C_g,|C_s\Gamma_2| \lesssim 10^{-2}$), all binaries have high eccentricities, $e_b > 0.5$, and both GDF and SDF lead to bimodal eccentricity distributions. Eccentricity appears to be a strong function of impact parameter --- each distinct region of Jacobi capture leads to a different eccentricity distribution (Figure \ref{fig:eb_plots}c). In the high-friction regime, under GDF ($C_g > 10^{-2}$) the range of possible $e_b$ broadens drastically --- for $C_g \gtrsim 10^{-1}$, we find a range of $1-e_b \in [7\times10^{-6},0.993]$ --- and the median shifts to  higher $e_b$ with higher $C_g$. Under SDF, there is no such development --- eccentricities rarely exceed $e_b = 0.9$. 

From Figure \ref{fig:eb_plots}d, we comment that the impact parameters yielding the highest-$e_b$ binaries (dark purple dots) roughly coincide with the transitions in impact parameter space between yielding prograde captures and regions yielding retrograde captures. (In this paper, we do not show the dependence of orientation on $\xi_0$, but our results are consistent with those shown in Figure 9 of \citet{Qian_2024}.) This result is similar to the trends in Figure 8 of \citet{Delaurentiis_2023}. The systems at these transitional regions undergo very close encounters, which apparently result in high eccentricites, but which could also lead to GW-assisted capture and inspiral on short timescales (see, e.g., discussion in \S6 of \citealp{Qian_2024}).

Finally, we examine the inclinations of newly formed BBHs in Figure \ref{fig:inc_hists}. All interactions with $f_I = 0$ yield binaries that are perfectly aligned (prograde) or anti-aligned (retrograde) with the reference orbit around the SMBH (i.e., the specific angular momentum vector is parallel or anti-parallel to $\hat{\pmb{\zeta}}$); however, in a real disk of single BHs with non-zero thickness, one would make BBHs with a continuous range of inclinations relative to the collective normal vector of the disk. We calculate this inclination, $\theta_b$, for all captured binaries from our initially inclined interactions. The resultant distribution is shown in Figure \ref{fig:inc_hists}.

\section{Discussion}\label{sec:discussion}

\subsection{GDF in the Context of Recent Work}\label{sec:compare}
\begin{figure*}
    \centering
    \includegraphics[width=.8\textwidth]{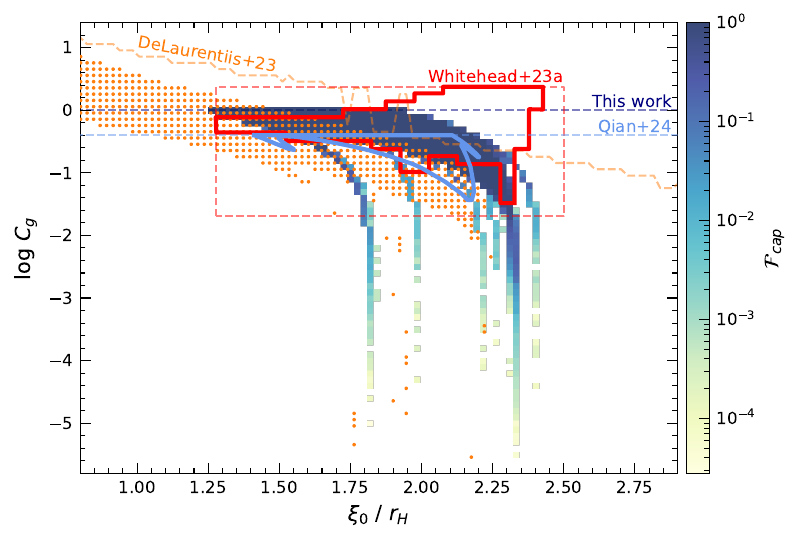}
    \caption{Regions of impact parameter ($\xi_0$) -- friction coefficient ($C_g$) space yielding binary formation in recent studies of gas-driven BBH formation. The capture regions reported in \citet{Qian_2024} and \citet{Whitehead_2023} are shown as solid blue and red outlines, respectively. Individual test cases leading to capture from \citet{Delaurentiis_2023} are shown as orange points. Our capture histogram for class (1) interactions under GDF (left panel, Figure \ref{fig:2d_hists}) is shown behind the other data --- the colorbar on the right applies to our data only. Dashed lines demarcate regions probed by various works --- \citet{Whitehead_2023} sample inside their rectangle, while \citet{Delaurentiis_2023}, \citet{Qian_2024}, and this work sample below their respective dashed lines. These three works are based on analytical prescriptions for GDF using formulae from \citet{Ostriker_1999}, while \citet{Whitehead_2023} used hydrodynamic simulations.}
    \label{fig:compare_1}
\end{figure*}
\begin{figure*}
    \centering
    \includegraphics[width=\textwidth]{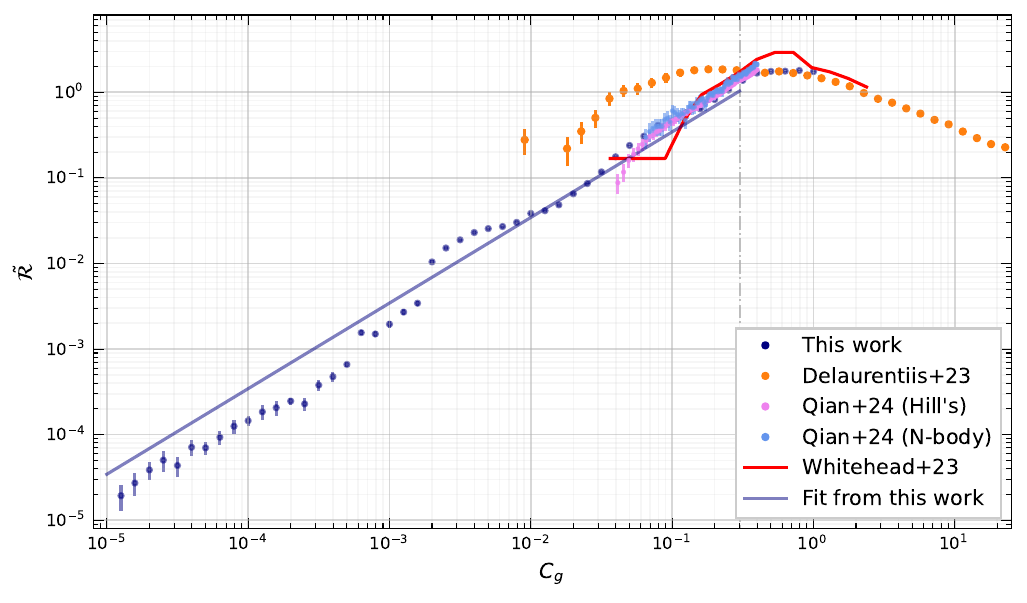}
    \caption{Estimates of the scaled binary formation rate $\rate$ (eq.\ \ref{eq:rate_scale}) as a function of friction coefficient $C_g$, using data from several of recent studies gas-driven BBH formation. The rates are consistent between most (but not all) studies where they overlap, mostly in the limited range $C_g \in [0.03,0.3]$. The vertical, dot-dashed line denotes the coefficient at which frictional forces are comparable to mutual gravitational forces at the Hill radius  --- here we expect a change in behavior of the capture rate (see \S\ref{sec:preint_effects}). The solid blue line represents our best-fit approximation to the capture rate, equation (\ref{eq:bestfit}). Note the variety of methods represented in this figure: this work and \citet[][Hill's]{Qian_2024} used Hill's problem with an analytic prescription for GDF, \citet{Delaurentiis_2023} and \citet[][N-body]{Qian_2024} used N-body simulations with similar analytic prescriptions, and \citet{Whitehead_2023} used hydrodynamic Hill's problem simulations. To avoid confusion due to data points with large error bars, we only show estimated rates when the rate is based on at least two captures.}
    \label{fig:compare_2}
\end{figure*}

In Figures \ref{fig:compare_1} and \ref{fig:compare_2}, we compare our results to those of \citet{Delaurentiis_2023}, \citet{Qian_2024}, and \citet{Whitehead_2023}. In general, there is rough consistency between capture rates as a function of  $C_g$ between several of the studies, though there is some disagreement in the region of $\xi_0$--$C_g$ space leading to capture.

In this section we thoroughly discuss the three papers for which we show results, listed above; we then discuss several other relevant papers on this topic. 

\vspace{.2cm}
\noindent \textit{\citet{Delaurentiis_2023}} --- This paper studies  binary formation through three-body simulations. Their GDF formula is that of \citet{Ostriker_1999} including the full dependence on Mach number (recall that we have assumed a Mach number $\ll 1$). They explore linearly spaced impact parameters $\xi_0 \in [0.5,2.5]$ (labeled $b$ in their notation). They also test a set of logarithmically spaced gas densities spanning 13 orders of magnitude, which, for their fiducial parameters, corresponds to a range of friction coefficients $C_g \in [7.19\times10^{-10},5.71\times10^3]$. (For reference, their Figure 10 spans $C_g \in 7.19\cdot[10^{-4},10^1]$.)

The regions of $\xi_0$--$C_g$ space in which we see the most captures are similar in shape to those found in \citet{Delaurentiis_2023}; however, their capture regions seem to be shifted to lower values of both GDF strength and impact parameter (see Figure \ref{fig:compare_1}). For $C_g \lesssim 0.3$, the capture rates calculated based on the \citet{Delaurentiis_2023} data appear to be inflated by a factor $\approx 2$--$5$ relative to the capture rates from our work (see Figure \ref{fig:compare_2}). 

One possible reason for these discrepancies is that \citeauthor{Delaurentiis_2023}\ model GDF using the velocity-dependent model of \citet{Ostriker_1999}, whereas we take the subsonic limit of this prescription. Another possible reason is a difference in initial azimuthal separation --- \citeauthor{Delaurentiis_2023}\ initialize their interactions as perfectly circular, nearby orbits beginning with an azimuthal separation $10r_H$, while we start ours at $20r_H$. We worry that non-circular motion due to the gravitational influence of one body on the other may be non-negligible when the bodies are as close as $10r_H$, and as a result, the circular orbit ICs may be inaccurate. 

\vspace{.2cm}
\noindent \textit{\citet{Qian_2024}} --- This paper explores binary formation in an AGN disk through both three-body simulations and orbit integrations in Hill's problem. They use two models for friction: (\textit{i}) a GDF force linearly dependent on the non-Keplerian velocity, characterized by a constant timescale $\tau$ (their equation 4 --- this is the same model used in the present paper); and (\textit{ii}) a GDF force including the full velocity dependence of the \citet{Ostriker_1999} prescription, which is scaled by some constant $\tau_0$. In their Hill's problem (``sheared sheet'') integrations, they apply model (\textit{i}) with a dimensionless timescale $\hat{\tau} = \tau \Omega_K$. The authors also incorporate GW emission in a final set of integrations, which serves to increase the capture rates among encounters in which the separation becomes very small. 

Our dimensionless friction coefficient $C_g$ is related to the \citeauthor{Qian_2024}\ friction timescales as follows: their $(\tau_0\Omega_K)^{-1} = 3C_g$ (the factor of 3 is introduced through our taking the subsonic limit), and their $\hat{\tau}^{-1} = C_g$. Note that their $K = \xi_0/r_H$. 

We compare our results primarily to their simulations with friction model (\textit{i}), as this is equivalent to ours (their results are largely consistent across models and integration techniques).\footnote{While \citet{Qian_2024} appear to find captures at smaller impact parameters in their three-body simulations, these captures generally occur after two or more interactions separated by a synodic period, so the impact parameters during the interactions actually leading to capture are larger.} 
For their Hill's problem integrations, \citeauthor{Qian_2024}\ simulate interactions across a linearly spaced grid of impact parameters $K\in [1,2.5]$ and friction timescales $\hat{\tau}^{-1} \in [0.0,0.4]$. One of their findings is that prograde binaries cannot form with friction timescales $\hat{\tau}^{-1} \lesssim 0.1$ and retrograde binaries cannot form with $\hat{\tau}^{-1} \lesssim 0.03$. In contrast, we find captures for all values of $\hat\tau$, although the capture rate decreases as $\hat\tau$ increases. We believe that the reason for this difference is that we sample more finely in impact parameter. That said, for $\hat{\tau}^{-1} = C_g \gtrsim 0.03$, our overall capture rate is consistent with theirs (see Figure \ref{fig:compare_2}), and we are generally in agreement with the trends they find in the relative efficiency of retrograde and prograde capture. 

As in our comparison to \citet{Delaurentiis_2023}, the \textit{shapes} of the main capture regions in $\xi_0$--$C_g$ space (their $K$--$1/\hat{\tau}$ space) are remarkably similar, but our capture region is shifted to slightly larger impact parameters (see Figure \ref{fig:compare_1}). 

Lastly, while we do not show them here, we emphasize that \citeauthor{Qian_2024}\ find capture rates that are typically slightly \emph{lower} under their friction model (\textit{ii}) --- that using the full velocity dependence of the \citet{Ostriker_1999} prescription --- than under the model equivalent to the present work.

As the \citeauthor{Qian_2024}\ model (\textit{ii}) is equivalent to the \citet{Delaurentiis_2023} friction model --- with the exception of different initial conditions --- this suggests that the fact that our capture rates are lower than those of \citeauthor{Delaurentiis_2023}\ at the same $C_g$ is not solely the result of the difference in friction models used.

\vspace{.2cm}
\noindent\textit{\citet{Whitehead_2023}} --- This paper studies binary formation through two-dimensional hydrodynamic simulations of Hill's problem. They are unable to cover as broad (or as finely sampled) a parameter space as the semi-analytic works above, although they vastly expanded on the coverage of previous hydrodynamic simulations. They report results from a grid of linearly spaced impact parameters $b/r_H = \xi_0/r_H \in [1.3,2.5]$ and fiducial parameters that correspond to $C_g \in 2.01\cdot[10^{-2}, 10^{0}]$. 

The capture region shown in their Figure 10 is broadly similar to ours, as illustrated in Figure \ref{fig:compare_1}. This is an encouraging sign that analytic GDF prescriptions can reproduce some of the results of more realistic hydrodynamical simulations. However, \citeauthor{Whitehead_2023}\ find captures in a region centered around $\xi_0/r_H\simeq 2.2$ and $C_g\simeq 1$ where we do not, and we find captures at $C_g\lesssim 10^{-1.5}$, where they do not.  The latter discrepancy might be due to sparse sampling in their simulations, but the former may suggest that analytic prescriptions do not adequately describe the GDF-driven formation process in the strong friction regime ($C_g \gtrsim 1$; see also discussion of \citealp{Rowan_2023} below). In order to probe the lower $C_g$ regime, it may be useful for future hydrodynamic simulations to finely sample from impact parameters in regions of prevalent BBH formation indicated by our Figure \ref{fig:gdf_zoomed}. 

\vspace{.2cm}
\textit{Other notable works} --- To our knowledge, \citet{Li_2023} presented the first hydrodynamic simulations of GDF-driven binary formation in the context of BBHs in AGNi. The authors tested a range of disk masses and initial azimuthal separations, holding all other parameters constant (including impact parameter). They found that formation is possible when the gas density is high enough, but also that formation depends on the initial azimuthal separation (this is likely the result of interactions between each of the single BHs and surrounding gas before the two are strongly interacting with each other). The authors also found that, when the separation $\rho < r_H$, most energy dissipation occurs while the two BHs are moving away from each other (see their Figure 4). We do not find the same effect in our integrations --- dissipation of $E_{\text{H}}$ occurs roughly evenly over a cycle of subsequent pericenter passages --- which we suspect is a shortcoming of our GDF prescription. Finally, \citeauthor{Li_2023}\ reported an empirical criterion for binary formation in terms of the BH masses, the surface density, the distance to the SMBH, and the energy at first close approach (their equation 7). Unfortunately, we cannot easily re-cast this criterion in terms of $C_g$, so we do not directly compare our results.

The 3D hydrodyamic simulations of \citet{Rowan_2023} --- which preceded the 2D, Hill's problem hydrodynamic simulations of \citet{Whitehead_2023} from largely the same group of authors --- show that binary formation is also possible at much larger values of the friction coefficient ($C_g\gtrsim 100$). Their results are important in interpreting those from analytic GDF prescriptions (\citealp{Delaurentiis_2023}, this work) in the $C_g \gtrsim 1$ regime. There, simulations using semi-analytic prescriptions tend to find that the impact parameters yielding binary formation are smaller under higher $C_g$ (e.g., Figure \ref{fig:compare_1}); \citet{Rowan_2023} find binary formation at impact parameters between $\sim2$--$3$ Hill radii, under friction coefficients $C_g \sim 10^2$--$10^3$. This suggests either that the trend from analytic prescription results is reversed at even higher $C_g$ or that such prescriptions are not reliable at such high $C_g$. Considering this alongside the discrepancy between the capture regions from \citet{Whitehead_2023} and the present work at $C_g \simeq 1$, we support the conclusion from \citet{Rowan_2023} that simulations using semi-analytic GDF prescriptions differ fundamentally from hydrodynamical simulations in the strong-friction regime (which we argue starts at $C_g \approx 0.3$).

\citet{Rowan_2024} sampled more finely in impact parameter than \citet{Rowan_2023}, allowing them to arrive at criteria for BBH formation based on the binary energy at $\rho = 2r_H$ and radial separation at $\rho=r_H$. Such criteria allow one to estimate the chance of BBH formation from an interaction between two single BHs without integrating past the first pericenter passage. The \citet{Rowan_2024} criteria are generally in agreement with similar criteria given in \citet{Whitehead_2023}; the criteria from these companion papers represent the current state-of-the-art 
for estimating the probability of two bodies forming a binary under GDF \textit{before} the bodies interact strongly with each other.

\citet{Rowan_2023}, \citet{Rowan_2024}, and \citet{Whitehead_2023b} have also relaxed certain approximations made in the other GDF studies discussed here. For example, \citet{Rowan_2023} and \citet{Rowan_2024} showed that considering accretion onto the individual BHs during an interaction does not substantially impact the binary formation process, at least in the high-$C_g$ regime they probe \citep[see also][]{Suzuguchi_2024}. \citet{Whitehead_2023b} relaxed the assumption of isothermality in the gas --- i.e., they allowed the surrounding disk material to respond to heating from orbital energy dissipation as an adiabatic gas/radiation mixture. They found that binary formation is still possible, although the gas density in the immediate vicinity of the pair is substantially lowered relative to the initial gas density after the first few pericenter passages. Note that \citet{Delaurentiis_2023} tested some of their high-gas-density interactions by repeating integrations with GDF turned off after the first interaction --- this process seems to approximate the process observed by \citet{Whitehead_2023b}, and this still yielded binary formation in many cases.

Lastly, returning to works taking a semi-analytical approach, \citet{Rozner_2023} studied how the velocity with which two bodies enter their Hill radius impacts their chance of capture under GDF. Considering the completely general environment of two bodies interacting within a gaseous medium, they present criteria for capture based on this velocity and local gas conditions, as well as a formula for capture rate. This analysis is of the two-body problem, i.e., there is no central mass. With these assumptions, they find that capture should be inefficient in AGN disks unless the disk is geometrically thin. We agree that disks with smaller scale heights (all else equal) will yield more efficient binary formation, as we find $\rate \propto C_g \propto (h/r)^{-3}$; however, this expected dependence on $h/r$ is not nearly as strong as what they find (e.g., see their Figure 11, though note that this presents capture rates integrated over an entire disk while our $\rate$ is local). We stress that \citeauthor{Rozner_2023}\ only consider one possible orientation of the velocity vector upon entering the Hill radius and that, as noted above, their capture rate equation neglects the shearing effects of a central mass. In general, this setup is too dissimilar to ours to make a direct comparison. 

\vspace{.2 cm}
\textit{Conclusions from comparison} --- We suggest that one can estimate the rate of binary formation per unit area, given a local number density of single BHs and a set of disk properties, by evaluating $C_g$ for that set of properties and (if $C_g \lesssim 0.3$) using the simple linear relationship $\rate = (3.4\pm0.3)\,C_g$ to find the corresponding formation rate. For stronger friction ($C_g \gg 0.3$), there are discrepancies in conditions leading to capture between results from semi-analytical GDF prescriptions \citep[e.g.,][]{Delaurentiis_2023} and more rigorous hydrodynamical results \citep[e.g.,][]{Rowan_2023}, so we do not provide such a simple prescription for formation rate in this regime. 

First, we point out that $C_g$ is linearly proportional to the mass $m$ of the involved single BHs. From the above conclusion that $\rate \propto C_g$ for $C_g \lesssim 0.3$, it would seem that among a population of single BHs with a range of masses but similar orbits (i.e., same eccentricity, inclination) at a given location in an AGN disk, the more massive BHs will form binaries by GDF-driven dissipation more frequently ($\rate \propto m$).

Under weak friction ($C_g \lesssim 0.1$), binaries formed by this mechanism are typically retrograde with respect to the disk angular momentum; under stronger friction ($0.1 < C_g \lesssim 1$), prograde captures and retrograde captures occur at similar rates. These results are supported by studies using both semi-analytic prescriptions (this work, \citealp{Qian_2024}) and hydrodynamical simulations \citep{Whitehead_2023}. Under much stronger friction ($C_g \sim 100$), this formation mechanism preferentially yields retrograde binaries  (\citealp{Rowan_2024}; hydrodynamical simulations). A binary formed under this mechanism that is \textit{retrograde} relative to the accretion disk would have \textit{negative} effective spin $\chi_{\text{eff}}$ if (\textit{i}) the single BHs began with prograde spins relative to the disk and (\textit{ii}) the spins did not change significantly relative to an inertial frame over the course of the binary formation process. The validity of these points is poorly assessed, so we cannot make a confident statement on the expected $\chi_{\text{eff}}$ distribution of BBH mergers from this binary formation mechanism.

None of the aforementioned studies --- nor this one --- have examined the dependence of formation rate on the mass \textit{ratio} of the interacting single BHs, yet any strong trends in BBH formation rate as a function of mass ratio would be of great observational interest. Based on our \S\ref{sec:results}, it seems that for $C_g \lesssim 10^{-2}$, long-term binary formation occurs among the initial conditions that would typically yield Jacobi capture in a friction-free model (this holds for both circular/flat and eccentric/inclined initial orbits). Therefore it may be possible to gain insights into the rate of GDF-driven BBH formation for various mass ratios by studying the influence of mass ratio on the regions of parameter space leading to prevalent Jacobi capture under no friction. Such work would need to cover one axis of parameter space fewer than repeating something like the present study with different mass ratios, as it would only consider $C_g=0$.

\subsection{Realistic Friction Strengths} \label{sec:typical_frics}

So far, we have presented BBH formation rates for arbitrary values of $C_g$ and $C_s$. In this subsection, we discuss expectations for the relative importance of GDF and SDF in forming binaries in galactic nuclei, and we provide estimates for typical values of $C_g$ and $C_s$. 

\subsubsection{Comparing GDF and SDF} \label{sec:Cg_over_Cs}

We are interested in the relative strength of GDF and SDF in a given galactic nucleus. Dividing equation (\ref{eq:K}) by equation  (\ref{eq:C}) and taking $m(m+m_a) \approx m^2$ (i.e., the mass $m$ BHs are much more massive than field stars --- mass $m_a$), one finds
\begin{equation}\label{eq:K_over_C}
    \frac{C_g}{C_s} = \frac{1}{3}\frac{\rho_g}{\rho_s}\left(\frac{r}{h}\right)^3 \frac{1}{\ln \Lambda}. 
\end{equation}
This is approximately the ratio of the densities of the media contributing the two types of dynamical friction times the ratio of their velocity dispersions, cubed. 

In general, if an AGN accretion disk is present, standard disk models \citep{Sirko_2003,Thompson2005} suggest the ratio $C_g/C_s$ will be large --- that is, they suggest that GDF will be significantly more important than SDF. To illustrate this, we first assume $\Lambda \approx N_\star$, the number of stars in the NSC. We make the very rough approximation that $N_\star \sim 10^7$ for a NSC of mass $M_{\text{NSC}}\approx$ a few$\,\cdot\,10^7M_\odot$ \citep[see review by][]{NSC_review_2020}. Then, generously, $3\ln\Lambda \sim 10^2$. In the fiducial setups shown in Figure 6 of \citet{Thompson2005} and Figure 4 of \citet{Tagawa_2020}, the scale height $(h/r) \sim 10^{-3}$ for AGN disks at scales $\sim$ 1 pc. Using these values, we can estimate
\begin{equation}\label{eq:K_over_C_scaling}
    \frac{C_g}{C_s} \sim 10^7 \left(\frac{\rho_g}{\rho_s}\right)\left(\frac{10^{-3}}{h/r}\right)^{3}\frac{10^2}{3\ln\Lambda}.
\end{equation}
In the same Figure 4 of \citet{Tagawa_2020}, $\rho_g/\rho_s$ is of order a few at 1 pc from a SMBH, so $C_g/C_s \gg 1$. With these values, then, we expect GDF to dominate the BBH formation rate at these distances from a SMBH when a gas disk is present. 

That said, recent results from cosmological simulations \citep{Hopkins_2024a,Hopkins_2024b} suggest that AGN disks are magnetically dominated; thus their structure appears vastly different from the models mentioned above (which assumed a geometrically thin \citet{Shakura_1973} disk in the inner regions, combined with a thermally supported outer region). This finding inspired the derivation of a new model for SMBH accretion disks \citep{Hopkins_2024c}, which typically exhibits much lower gas densities and larger scale heights.  In this case the ratio $C_g/C_s$ would be considerably lower. We make no statements on how this change would impact, e.g., the tendency of the orbits of single BHs to become aligned with AGN disks or the efficacy of in-disk migration traps (arguments 2 and 3 of \S\ref{sec:state}) but if these processes still work the importance of SDF relative to GDF in forming BHBs would be much higher. 

Lastly, we emphasize that a typical galaxy spends only a small fraction of its lifetime as an AGN ($\sim10^7$--$10^9$ yr; e.g., \citealp{Yu_2002,Marconi_2004}), and during the inactive phase SDF will continue to act while GDF will not. Thus even if $C_g/C_s$ is large, SDF could contribute an appreciable fraction of dynamically formed BBHs when integrating over the age of the Universe.

\subsubsection{Typical Values of $C_g$ and $C_s$}\label{sec:typical_vals}

AGN accretion disks are unresolved in almost all cases and their properties must be inferred from spectra and theoretical models. The properties of nuclear star clusters (NSCs) are also poorly known, in part because these objects are not well-resolved in most galaxies (\citealt{NSC_review_2020} provide a thorough review of the recent state of NSC observations). Within these limitations, we here attempt to give a sense of possible physical values of the friction coefficients $C_g$ and $C_s$. We take our single BH mass to be $m = 20M_\odot$. (We emphasize that both $C_g$ and $C_s$ are linearly proportional to $m$.)

\begin{figure}
    \centering
    \includegraphics[width=.47\textwidth]{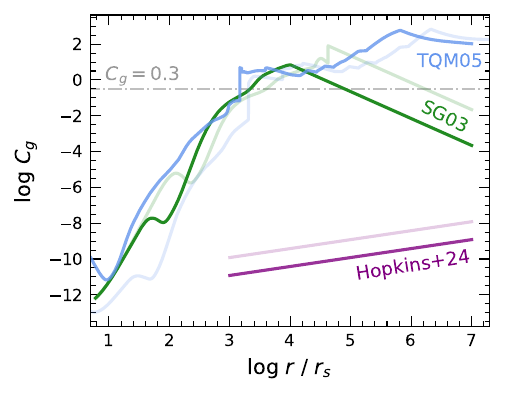}
    \caption{Values of the dimensionless gaseous friction coefficient $C_g$ as a function of radius $r$ for several models of AGN disks around $M_\bullet=10^6$ and $10^7M_\odot$ SMBHs (faint and dark lines, respectively). The horizontal axis is scaled by the SMBH Schwarzchild radius $r_s$, which is $\sim 10^{-7}$ and $10^{-6}$ pc for $10^6$ and $10^7M_\odot$ SMBHs, respectively. These profiles illustrate the uncertainties and trends in the values of $C_g$ for real disks. The friction coefficient varies by many orders of magnitude as a function of radius, and between the displayed models --- like \citet[SG03]{Sirko_2003}, \citet[TQM05]{Thompson2005} and  \citet[Hopkins+24]{Hopkins_2024c}. We show a horizontal line at $C_g = 0.3$ --- below this, the BBH formation rate is roughly linearly proportional to $C_g$ (eq.\ \ref{eq:bestfit}), while for larger values of $C_g$ the assumptions of our analytic GDF models may not be valid. The SG03 and TQM05 profiles were created with \texttt{pagn} \citep{pagn_code}.}
    \label{fig:cg_profiles}
\end{figure}

To estimate the gas dynamical friction coefficient $C_g$, we look to standard models of AGN disks in the literature. These are typically those of \citet{Sirko_2003} and \citet{Thompson2005}. Very recently, \citet{Gangardt_2024} presented the open-source code \texttt{pagn} \citep{pagn_code}, with which one can easily create one-dimensional AGN disk models based on the \citet{Sirko_2003} and \citet{Thompson2005} prescriptions using up-to-date opacity tables. We use \texttt{pagn} to create two simple examples of disk profiles around $M_\bullet = 10^6$ and $10^7M_\odot$ SMBHs. The \citet[SG03]{Sirko_2003} disk has a radiative efficiency $\epsilon_S = 0.1$; an Eddington luminosity ratio $L/L_{\text{Edd}} = 0.5$, where $L_{\text{Edd}}$ is the Eddington luminosity; and a Shakura-Sunyaev $\alpha$ viscosity with $\alpha = 0.1$ and viscosity proportional to the total pressure. The \citet[TQM05]{Thompson2005} model has a prescribed outer radius at $10^7r_s$, where $r_s = 2\mathbb{G}M_\bullet/c^2$ is the Schwarzschild radius of the SMBH; star formation efficiency $\epsilon_T = 10^{-3}$; supernova radiative efficiency $\xi=1$; torque efficiency $m_T=0.2$; and an outer accretion rate $\dot{M}_{\text{out}} = 1.5\times10^{-2}M_\odot$ yr$^{-1}$. (See \citealp{Sirko_2003}, \citealp{Thompson2005}, or the overview of both models in \citealp{Gangardt_2024} for more on all of these parameters.) These models are shown in Figure \ref{fig:cg_profiles}. 

Finally, we create a profile of $C_g$ based on the prescription in \citet{Hopkins_2024c}, which describes a magnetically dominated disk based on the results of cosmological zoom-in simulations. Using their equations (5) and (6), we can arrive at radial profiles of $\rho_g$ and $h/r$ --- and thus a profile of $C_g$ --- by setting three parameters: an SMBH mass $M_\bullet = 10^6$ or $10^7M_\odot$, a free-fall radius $r_{\text{ff}} = 5$ pc, and an Eddington ratio $\dot{m} \equiv \dot{M}/\dot{M}_{\text{Edd}} = 0.5$; finally, we assume the orbital frequency is Keplerian. These choices yield the profiles of $C_g$ shown in Figure \ref{fig:cg_profiles}. We only show this profile down to $r/r_s = 10^3$, which is roughly the innermost radius for which their cosmological simulations tracked disk properties.

The models shown here are mainly intended to illustrate that the predicted friction coefficients from different models can vary by several orders of magnitude at the same radius, even for an SMBH with a fixed mass and accretion rate. This variation is most dramatic between the \citet{Hopkins_2024c} model for a magnetically dominated AGN disk and the gas or radiation pressure dominated disks of \citet{Sirko_2003} and \citet{Thompson2005}. Thus the approximations made in developing our GDF prescription, and  our conclusion that the formation rate $\rate \propto C_g$, contribute negligible errors in estimating BBH formation rates compared to our uncertainty regarding the structure of AGN disks.

To provide examples of the stellar dynamical friction coefficient $C_s$, we use the nearby NSCs in the Milky Way (MW) and the Andromeda galaxy (M31). We set the typical field star mass $m_a = 1M_\odot$. For the MW, the mass of the central black hole is $M_\bullet = 10^{6.6}M_\odot$ \citep{NSC_review_2020}; the stellar mass within $r = 1$ pc is $M_{\text{NSC}} \simeq 10^{6.6}M_\odot$, and the mass density there is $\rho_s(1\mbox{\,pc}) \simeq 10^{5.3}M_{\odot}$ pc$^{-3}$ \citep{Schodel_2007}. Based on these values, we assume that the star cluster contains $N_\star \approx 10^{6.6}$ stars and set the Coulomb logarithm $\ln\Lambda = \ln N_\star$. Finally, we assume the logarithmic density slope $\gamma = 1.5$ \citep[e.g.,][]{NSC_review_2020}. Our results are not sensitive to the value of $\gamma$ in the range  $[1.5,2.5]$. From equation (\ref{eq:C}), these values yield $C_{s,\,\text{MW}} \sim 10^{-4}$. For M31, we have $M_\bullet \simeq 10^{8.2}M_\odot$ and $M_{\text{NSC}} \simeq 10^{7.7}M_\odot$ \citep{NSC_review_2020}, and based on the profiles of \citet{Pechetti_2020}, $\rho_s(1\mbox{\,pc}) \sim 10^5M_{\odot}$ pc$^{-3}$. The same calculation that we carried out for the MW yields  $C_{s,\,\text{M31}} \sim 10^{-7}$. For $\gamma = 1.5$, $\Gamma_2 = -0.06$, so we have $|C_s\Gamma_2|_{\text{MW}} \sim 6\times10^{-5}$ and $|C_s\Gamma_2|_{\text{M31}} \sim 6\times10^{-9}$.

Comparing these values to the $C_g$ profiles in Figure \ref{fig:cg_profiles}, we can say that GDF is typically more important than SDF in the presence of a \citet{Sirko_2003} or \citet{Thompson2005} AGN disk. SDF and GDF likely play a comparable role if the \citet{Hopkins_2024c} disk model is more appropriate, and as discussed in the previous subsection, SDF is always more important than GDF during the inactive phase of a galactic nucleus.

\section{Conclusions}
The main goal of this paper has been to estimate the formation rate of binary black holes (BBHs) from populations of single BHs on nearly circular, nearly coplanar orbits around a (super)massive black hole. During close interactions between two single BHs on similar orbits, permanent capture is induced through energy dissipation via gas dynamical friction (GDF) or stellar dynamical friction (SDF), applicable within a gas disk or stellar cluster, respectively. We make numerical estimates of formation rates from this process by simulating gravitational interactions between two single BHs using Hill's approximations. The strength of the friction is parametrized by the dimensionless friction coefficients $C_g$ or $C_s$, which are related to the conditions in a gas disk (gas density and sound speed) or stellar cluster (stellar density and velocity dispersion) by equations (\ref{eq:K}) and (\ref{eq:C}). These are derived from the \citet{Ostriker_1999} and \citet{Chandrasekhar_1943} prescriptions for GDF and SDF, respectively. 

We find that the binary formation rate per unit area $\mathcal{R}$ (eq.\ \ref{eq:caprate}; \citealp{Qian_2024}) scales roughly linearly with both $C_g$ and $C_s$. That is, when only considering circular and flat orbits, 
\begin{equation}\label{eq:final_scaling_Cg}
    \frac{\mathcal{R}_{\text{GDF}}}{N^2 \Omega_K R_H^2} \simeq 3.5\pm0.3 \left[\frac{4\pi}{3}\frac{1}{(h/r)^3}\frac{\rho_g m r^3}{M_{\bullet}^2}\right]
\end{equation}
and 
\begin{equation}\label{eq:final_scaling_Cs}
    \frac{\mathcal{R}_{\text{SDF}}}{N^2 \Omega_K R_H^2} \simeq 4.4\pm0.5 \left[4\pi \ln\Lambda \frac{\rho_s m r^3}{M_{\bullet}^2} \right]|\Gamma_2|,
\end{equation}
where the prefactors are best-fit parameters based on the Monte Carlo integrations conducted in this work, and the term in square brackets in equations (\ref{eq:final_scaling_Cg}) and (\ref{eq:final_scaling_Cs}) is $C_g$ and $C_s$, respectively. In these equations, $m$ is the mass of the BHs (assumed equal); $r$ the distance from the SMBH of mass $M_\bullet$; $\rho_g$ and $\rho_s$ the local mass densities of gas and stars, respectively; $(h/r)$ the fractional scale height of the accretion disk; $\ln\Lambda$ the Coulomb logarithm; $\Gamma_2$ a function of $\gamma$, the density profile parameter ($\rho_s\sim r^{-\gamma}$) of the nuclear stellar cluster (eq.\ \ref{eq:gamma_2}); $N$ the surface number density of BHs on circular orbits in the midplane of the disk; $\Omega_K$ the local Keplerian angular speed; and $R_H$ the mutual Hill radius of two BHs of mass $m$ at $r$ (eq.\ \ref{eq:hilldef}).

These scalings are consistent with our data for friction coefficients $C_g \lesssim 0.3$ and $|C_s\Gamma_2| \lesssim 0.01$. For 
GDF, this upper limit occurs roughly at the value of $C_g$ at which friction begins to have a significant effect on the interaction before the bodies enter their mutual Hill radius (see \S\ref{sec:preint_effects}) and thus probably represents a physical change in the dynamics of the encounter. For SDF, this limit occurs roughly when we start to see substantial differential radial drift before the bodies strongly interact, so our model in which the two bodies approach each other on circular orbits is oversimplified. 

Our capture rate results for GDF are largely consistent with recent works in the literature (see Fig.\ \ref{fig:compare_2}), though the details of the  distribution of captures in impact parameter--friction coefficient space vary somewhat between investigations (see \S\ref{sec:compare} and Fig.\ \ref{fig:compare_1}).

We also present the first estimates of the capture rate between two BHs on initially eccentric or inclined orbits. In both cases, the rates are similar to those for BHs on circular/flat orbits when the eccentricity or inclination is $ \lesssim R_H/\bar{a}$ (where $\bar{a}$ is the distance to the central mass in Hill's problem and $R_H$ is the mutual Hill radius). For larger eccentricities or inclinations the capture rate falls, slowly for eccentricities and more rapidly for inclinations (see Fig.\ \ref{fig:rate_ei}). It would be worthwhile to determine the steady-state distribution of single BH eccentricities and inclinations within an AGN disk at a given distance from a SMBH as input for estimating the overall binary formation rate (see eq.\ \ref{eq:caprate} and \S\ref{sec:calculate} more generally).

In the disks of AGNi, we generally expect GDF to dominate SDF in determining the formation rate of BBHs (see \S\ref{sec:Cg_over_Cs}, and especially eq.\ \ref{eq:K_over_C_scaling}). However, (\textit{i}) the expected capture rates due to GDF are uncertain by many orders of magnitude (see Fig.\ \ref{fig:cg_profiles}); (\textit{ii})  AGNi are short-lived, whereas SDF acts at all times; SDF could have a significant effect if a disk of single BHs were formed during an AGN state and their velocity dispersion remained low following the dissolution of the accretion disk. Therefore SDF from the nuclear star cluster cannot always be neglected when considering dynamical formation of BBHs in galactic nuclei.

Alhough we have focused on galactic nuclei, the results of our work can be applied to other systems. On planetary system scales, massive (but not gap-clearing) planetary cores interacting within a gaseous protoplanetary disk would experience GDF that might lead to binary formation. Even once the gas in the disk has disappeared, a disk of smaller bodies would exert dynamical friction analogous to GDF on larger planetesimals. (In this way, our work is a development on \citealt{Goldreich_2002}; our $C_g$ is related to their drag term $D$, and our result that $\rate \propto C_g$ down to at least $C_g \sim 10^{-5}$ suggests that the trend in their $L^2$ channel continues for drag strengths several orders of magnitude lower than they studied.)

Our prescription for binary formation by SDF might also be relevant for globular clusters, if some or all of the stars in these clusters are formed in or dragged into thin gaseous disks. 

Finally, we stress that our results are based on several assumptions and approximations --- the relative speed of the BHs is assumed to be much less than the sound speed in the disks; the wakes of the BHs are assumed to not interact; the BH masses are set to be equal and assumed to be much greater than the masses of the stars in the nuclear star cluster (SDF); the spatial excursions of the BH orbits are assumed to be not much larger than the Hill radius; etc. All of these approximations are discussed in the text; they likely add errors to our results, but in systems for which our assumptions are valid these errors are small compared to the uncertainties in the typical parameters of AGN disks and galactic nuclei. Furthermore, the approximate agreement between our GDF capture rates, those from other recent works using analytic GDF prescriptions \citep[specifically][]{Delaurentiis_2023,Qian_2024}, and those from the hydrodynamic simulations of \citet{Whitehead_2023} provide some assurance that our model captures the main features of BBH formation by dynamical friction.

\software{
Python3 \citep{python}, 
scipy \citep{scipy},
numpy \citep{numpy},
pandas \citep{pandas},
matplotlib \citep{matplotlib},
pagn \citep{pagn_code}
}

\vspace{.2 cm}
We thank Christopher Matzner, Norman Murray, Maya Fishbach, Aditya Vijaykumar, and Jiaru Li for thoughtful comments and conversations. We gratefully acknowledge Stanislav DeLaurentiis, Kecheng Qian, and Henry Whitehead for sharing data from their respective papers. This work was supported in part by the Natural Sciences and Engineering Research Council of Canada (NSERC), funding reference number RGPIN-2020-03885.

\bibliography{references}

\appendix
\section{Derivation of Stellar Dynamical Friction Equation}\label{sec:appendix}

From equations (7.92) and (8.7) of \citet{Binney_2008}, the dynamical friction force acting on a body of mass $m$ moving with velocity vector $\dot{\mathbf{r}}$ through a field of bodies with mass $m_a$ can be written 
\begin{equation}
    m\ddot{\mathbf{r}} = -\frac{4\pi\mathbb{G}^2\rho_s m(m+m_a)\ln\Lambda}{\sigma^2}G(X)\frac{\dot{\mathbf{r}}}{v},
\end{equation}
where $\mathbb{G}$ is the gravitational constant;
\begin{equation}
    G(X) \equiv \frac{1}{2X^2}\text{erf}(X) - \frac{1}{X\sqrt{\pi}}\exp{(-X^2)};
\end{equation}$X\equiv v/\sqrt{2}\sigma$, with $v=|\dot{\mathbf {r}}|$ and the one-dimensional velocity dispersion of the field stars $\sigma$; $\rho_s$ the local field density; and $\ln\Lambda$ the Coulomb logarithm, with $\Lambda$ defined in equation (7.84) of \citet{Binney_2008}. If we have two bodies with equal masses $m_1=m_2=m$, we can define some $\dot{\mathbf{s}}$ such that their individual velocity vectors are given by $\dot{\mathbf{r}}_2 = \dot{\mathbf{r}} - \dot{\mathbf{s}}$ and $\dot{\mathbf{r}}_1 = \dot{\mathbf{r}} + \dot{\mathbf{s}}$ (where $\dot{\mathbf{r}}$ is the center-of-mass velocity and $2\dot{\mathbf{s}}$ is the relative velocity vector). Then the differential acceleration on these bodies is
\begin{equation}\label{eq:base_diff}
    \ddot{\mathbf{r}}_2 - \ddot{\mathbf{r}}_1 = - \frac{4\pi\mathbb{G}^2\rho_s(m+m_a)\ln\Lambda}{\sigma^2}\mathbf{b},\quad \text{where} \quad \mathbf{b} \equiv G(X_2)\frac{\dot{\mathbf{r}}_2}{v_2} - G(X_1)\frac{\dot{\mathbf{r}}_1}{v_1},
\end{equation}
where we have abbreviated $X(v_n)$ as $X_n$.
For bodies on nearby, initially circular orbits, as considered here, the Keplerian velocity dominates the velocity vectors, and the separation of the bodies on their initial orbits is small enough compared to their distance from the central object that their Keplerian velocities are approximately equal; i.e., $\dot{\mathbf{r}}_1 \simeq \dot{\mathbf{r}}_2$. Therefore we assume that $|\dot{\mathbf{s}}| = |\dot{\mathbf{r}}_1 - \dot{\mathbf{r}}_2|/2 \ll |\dot{\mathbf{r}}|$ --- this inequality may be violated for short periods of time during close encounters between the two bodies, but this should not have a significant effect on our results.

When $\dot{\mathbf{s}} = 0$, the vector $\mathbf{b}=0$; we can Taylor expand around $\dot{\mathbf{s}} = 0$ to find the behavior of equation (\ref{eq:base_diff}) when $|\dot{\mathbf{s}}|$ is small. This expansion gives 
\begin{equation}
    \mathbf{b} \simeq \left.\frac{\partial \mathbf{b}}{\partial\dot{\mathbf{s}}}\right|_{\dot{\mathbf{s}} = 0} \dot{\mathbf{s}} + \mathcal{O}(|\dot{\mathbf{s}}|^2).
\end{equation}
The first-order term is a matrix $(\partial\mathbf{b}/\partial\dot{\mathbf{s}})$ right-multiplied by a vector $(\dot{\mathbf{s}})$; the $j$-th column of the matrix is given by
\begin{align}\label{eq:first-order}
    \frac{\partial}{\partial\dot{s}_j}\left[G_2\frac{\dot{\mathbf{r}}_{2}}{v_2} - G_1\frac{\dot{\mathbf{r}}_{1}}{v_1}\right] = 
    \frac{\partial G_2}{\partial X_2}&\frac{\partial X_2}{\partial\dot{s}_j}\frac{\dot{\mathbf{r}}_2}{v_2}
    + \frac{G_2}{v_2} \frac{\partial\dot{\mathbf{r}}_2}{\partial\dot{s}_j}
    - G_2 \frac{\partial v_2}{\partial \dot{s}_j}\frac{\dot{\mathbf{r}}_2}{v_2^2}
    \nonumber \\
    &- \frac{\partial G_1}{\partial X_1}\frac{\partial X_1}{\partial\dot{s}_j}\frac{\dot{\mathbf{r}}_1}{v_1}
    - \frac{G_1}{v_1} \frac{\partial\dot{\mathbf{r}}_1}{\partial\dot{s}_j}
    + G_1 \frac{\partial v_1}{\partial \dot{s}_j}\frac{\dot{\mathbf{r}}_1}{v_1^2},
\end{align}
where we have abbreviated $G(X_n)$ as $G_n$. Note that $\partial X_n / \partial \dot{s}_j = (\partial v_n/\partial \dot{s}_j) / \sqrt{2}\sigma$. Note also that
\begin{align}
    \frac{\partial\dot{\mathbf{r}}_1}{\partial\dot{s}_j} = \hat{\mathbf{e}}_j, 
    \quad\quad &\quad\text{   } \quad\quad 
    \frac{\partial v_1}{\partial \dot{s}_j} = \frac{\dot{s}_j + \dot{r}_{1,j}}{v_1},
    \nonumber \\ \label{eq:derivs}
    \frac{\partial\dot{\mathbf{r}}_2}{\partial\dot{s}_j} = -\hat{\mathbf{e}}_j, 
    \quad\quad &\text{and} \quad\quad 
    \frac{\partial v_2}{\partial \dot{s}_j} = \frac{\dot{s}_j - \dot{r}_{2,j}}{v_2}.
\end{align}
Plugging in these formulae, then evaluating the result at $\dot{\mathbf{s}} = 0$ (equivalently, $\dot{\mathbf{r}}_1 = \dot{\mathbf{r}}_2 = \dot{\mathbf{r}}$), equation (\ref{eq:first-order}) simplifies to 
\begin{equation}
   \frac{\partial}{\partial\dot{s}_j}
   \left[
   G_2\frac{\dot{\mathbf{r}}_{2}}{v_2} - G_1\frac{\dot{\mathbf{r}}_{1}}{v_1}
   \right] 
   = 
   -2\left(
   \frac{\partial G}{\partial X}\frac{\dot{r}_j\dot{\mathbf{r}}}{\sqrt{2}\sigma v^2} 
   - G\frac{\dot{r}_j\dot{\mathbf{r}}}{v^3} 
   + G\frac{\hat{\mathbf{e}}_j}{v}
   \right).
\end{equation}
In index notation, then, the first-order expansion is
\begin{equation}
    \mathbf{b} \simeq \frac{\partial}{\partial\dot{s}_j}\left[G_2\frac{\dot{r}_{2,i}}{v_2} - G_1\frac{\dot{{r}}_{1,i}}{v_1}\right] s_j 
    = 
    \left[
    \frac{\partial G}{\partial X}\frac{\dot{r}_i\dot{r}_j\hat{\mathbf{e}}_i}{\sqrt{2}\sigma v^2} 
    - G\frac{\dot{r}_i\dot{r}_j\hat{\mathbf{e}}_i}{v^3} 
    + G\frac{\hat{\mathbf{e}}_j}{v}
    \right]
    (\dot{\mathbf{r}}_2 - \dot{\mathbf{r}}_1)_j.
\end{equation}
We assume that all components of the velocity other than the local Keplerian velocity $v_K\hat{\mathbf{e}}_y$ are negligible, so to lowest order, $v_j\dot{\mathbf{r}} \simeq v_K^2\delta_{jy}\hat{\mathbf{e}}_y$ (and therefore $v \simeq v_K$). Then the expansion becomes
\begin{equation} \label{eq:full_expansion}
    \mathbf{b} \simeq
    \left[
    \frac{\partial G_K}{\partial X}\frac{\delta_{jy}\hat{\mathbf{e}}_y}{\sqrt{2}\sigma} 
    - G_K \frac{\delta_{jy}\hat{\mathbf{e}}_y}{v_K} 
    + G_K \frac{\hat{\mathbf{e}}_j}{v_K}
    \right]
    (\dot{\mathbf{r}}_2 - \dot{\mathbf{r}}_1)_j.
\end{equation}
Note $G_K=G(v_K/\sqrt{2}\sigma)$. The relevant derivative of $G(X)$ is
\begin{equation}
    \frac{\partial G}{\partial X} = 
    \left(
    \frac{2}{X^2\sqrt{\pi}} + \frac{2}{\sqrt{\pi}}
    \right)\exp(-X^2)
    - \frac{1}{X^3}\text{erf}(X).
\end{equation}
We can rewrite equation (\ref{eq:full_expansion}) as
\begin{equation}\label{eq:b_final}
    \mathbf{b} \simeq \frac{1}{v_K}\left[\Gamma^*_1(\dot{\mathbf{r}}_2 - \dot{\mathbf{r}}_1) + (\Gamma^*_2 - \Gamma^*_1)(v_{2y} - v_{1y})\hat{\mathbf{e}}_y\right],
\end{equation}
introducing the dimensionless constants $\Gamma^*_1$ and $\Gamma^*_2$, which are defined as (using $v_K / \sigma = \sqrt{2} X_K$)
\begin{align}
    \Gamma^*_1 \equiv G_K 
    &= 
    \frac{1}{2X_K^2}\text{erf}(X_K) - \frac{1}{X_K\sqrt{\pi}}\exp(-X_K^2)
    \\
    \Gamma^*_2 \equiv \frac{1}{\sqrt{2}\sigma}\frac{\partial G_K}{\partial X} 
    &= 
    -\frac{1}{X_K^2}\text{erf}(X_K) + \left(\frac{2}{X_K\sqrt{\pi}} + \frac{2X_K}{\sqrt{\pi}}\right)\exp(-X_K^2)
\end{align}
For a spherically symmetric and ergodic stellar system, following a density profile $\rho(r) \propto r^{-\gamma}$, one can solve the Jeans equations to obtain 
\begin{equation}
    \sigma = \frac{v_K}{\sqrt{\gamma + 1}} \quad\quad\text{and}\quad\quad X_K = \sqrt{\frac{\gamma + 1}{2}},
\end{equation}
so 
\begin{align}
    \Gamma^*_1 &= \frac{1}{\gamma + 1}\text{erf}\left(\sqrt{\frac{\gamma+1}{2}}\right) - \sqrt{\frac{2}{\pi(\gamma + 1)}}\exp\left(-\frac{\gamma+1}{2}\right)
    \\
    \Gamma^*_2 &= -\frac{2}{\gamma + 1}\text{erf}\left(\sqrt{\frac{\gamma+1}{2}}\right) + \left(2\sqrt{\frac{2}{\pi(\gamma + 1)}} + \sqrt{\frac{2(\gamma + 1)}{\pi}}\right)\exp\left(-\frac{\gamma+1}{2}\right)
\end{align}
Note that $\Gamma^*_1$ and $\Gamma^*_2$ depend only on the logarithmic density slope $\gamma$ of the star cluster. 
We now plug our expression for $\mathbf{b}$ (eq.\ \ref{eq:b_final}) back into equation (\ref{eq:base_diff}) to find
\begin{equation}
    \pmb{\rho}''_{\text{SDF}} = - C^*_s
    \left(\Gamma^*_1 \pmb{\rho}' + (\Gamma^*_2 - \Gamma^*_1)\eta' \hat{\pmb{\eta}}\right),
\end{equation}
where we have also converted to the dimensionless spatial and temporal coordinates $\pmb{\rho} = \mathbf{x}/3^{1/3}R_{\text{H}}$ and $t_d = \Omega t$ (with $R_H$ the mutual Hill radius and $\Omega$ the local Keplerian frequency).  The constant $C^*_s$ is given by
\begin{equation}
    C^*_s = \frac{4\pi\mathbb{G}^2\rho_s(m+m_a)\ln\Lambda}{\Omega v_K^3}(\gamma + 1) = 4\pi r^3\rho_s\frac{m+m_a}{M_{\bullet}^2}\ln\Lambda(\gamma + 1),
\end{equation}
where $M_\bullet$ is the mass of the central SMBH and $r$ is the distance from that to the bodies of interest. In the text, we define $C_s \equiv C^*_s / (\gamma + 1)$, $\Gamma_1 \equiv (\gamma + 1)\Gamma^*_1$, and $\Gamma_2 \equiv (\gamma + 1)\Gamma^*_2$ in order to make $C_s$ independent of $\gamma$. In keeping $C_s$ constant throughout a given interaction, we neglect any changes to the orbital radius $r$ over its duration.

\end{document}